\documentclass[useAMS,usenatbib]{mn2e}
\usepackage{times}
\usepackage{graphicx}
\usepackage{amsfonts} 
\def\ltsima{$\; \buildrel < \over \sim \;$}
\def\lsim{\lower.5ex\hbox{\ltsima}}
\def\gtsima{$\; \buildrel > \over \sim \;$}
\def\gsim{\lower.5ex\hbox{\gtsima}}
\def\ga{\mathrel{\hbox{\rlap{\hbox{\lower4pt\hbox{$\sim$}}}\hbox{$>$}}}}
\def\la{\mathrel{\hbox{\rlap{\hbox{\lower4pt\hbox{$\sim$}}}\hbox{$<$}}}}

\newcommand{\cMpch}{~h^{-1}~\mathrm{comoving~Mpc}}

\newcommand{\apj}{ApJ}
\newcommand{\apjl}{ApJL}
\newcommand{\apjs}{ApJS}
\newcommand{\aj}{AJ}
\newcommand{\mnras}{MNRAS}
\newcommand{\nat}{Nature}

\newcommand{\aap}{A\&A}

\newcommand{\prd}{PhRvD}
\newcommand{\physrep}{PhysRep}

\newcommand{\HI}{\rm H\, {\scriptstyle I} }

\newcommand{\MHI}{\mathrm{H\,  I } }

\title[Imaging Neutral Hydrogen During the EoR
with LOFAR]{Imaging neutral hydrogen on large-scales during the Epoch of Reionization
with LOFAR}

\author[Zaroubi et al.] {S. Zaroubi$^{1,2}$\thanks{E-mail:
saleem@astro.rug.nl}, A.G. de Bruyn$^{1,3}$, G. Harker$^{4}$, R. M. Thomas$^{5}$, P. Labropolous$^{1,3}$, \newauthor
V. Jeli\'c$^{3}$, L.V.E. Koopmans$^{1}$, M. A. Brentjens$^{3}$, G. Bernardi$^{6}$, B. Ciardi$^{7}$, S. Daiboo$^{1}$, \newauthor
 S. Kazemi$^{1}$, O. Martinez-Rubi$^{1}$, G. Mellema$^{8}$, A. R. Offringa$^{1}$, V.N. Pandey$^{1,3}$, J. Schaye$^{9}$,  \newauthor
V. Veligatla$^{1}$, H. Vedantham$^{1}$, S. Yatawatta$^{1,3}$
\\$^{1}$Kapteyn Astronomical Institute, University of Groningen, P.O. Box 800, 9700 AV Groningen,Tthe Netherlands
\\$^{2}$Department of Physics, The Technion, Haifa 32000, Israel
\\$^{3}$ ASTRON, P.O.Box 2, 7990 AA Dwingeloo, The Netherlands
\\$^{4}$ Center for Astrophysics and Space Astronomy, Dept. of Astrophysics and Planetary Sciences, University of Colorado at Boulder, CO 80309, USA 
\\$^{5}$ CITA, University of Toronto, 60 St George Street, M5S 3H8, Toronto, ON, Canada
\\$^{6}$ HarvardÐSmithsonian Center for Astrophysics, 60 Garden Street, Cambridge, MA 02138, USA
\\$^{7}$ Max-Planck Institute for Astrophysics, Karl-Schwarzschild-Stra?se 1, 85748 Garching, Germany
\\$^{8}$ Department of Astronomy and Oskar Klein Centre for Cosmoparticle Physics, AlbaNova, Stockholm University, SE-106 91 Stockholm, Sweden
\\$^{9}$ Leiden Observatory, Leiden University, PO Box 9513, 2300RA Leiden, The Netherlands
}

\begin{document}

\date{}

\maketitle

\begin{abstract}
The first generation of redshifted 21~cm detection experiments, carried out with arrays like LOFAR, MWA and GMRT, will have a very low signal-to-noise ratio per resolution element ($\lsim 0.2$). In addition, whereas the variance of the cosmological signal decreases on scales larger than the typical size of ionization bubbles, the variance of the formidable galactic foregrounds increases, making it hard to disentangle the two on such large scales. The poor sensitivity on small scales on the one hand, and the foregrounds effect on large scales on the other hand, make direct imaging of the Epoch of Reionization of the Universe very difficult, and detection of the signal therefore is expected to be statistical. 
Despite these hurdles, in this paper we argue that for many reionization scenarios low resolution images could be obtained from the expected 
data. This is because at the later stages of the process 
one still finds very large pockets of neutral regions in the IGM, reflecting the clustering of the large-scale structure, which stays strong up to scales of  
$\approx 120 \cMpch$ ($\approx 1^\circ$). The coherence of the emission on those scales allows us to reach sufficient S/N ($\gsim 3$)  so as to obtain 
reionization 21~cm images. Such images will be extremely valuable for answering many cosmological questions but above all they will be a very powerful 
tool to test our control of the systematics in the data. The existence of this typical scale  ($\approx 120 \cMpch$) also argues for 
designing future EoR experiments, e.g., with SKA, with a field of view of at least $4^\circ$.

\end{abstract}

\begin{keywords}
cosmology: theory Ð large-scale structure of Universe- observations  Ð diffuse radiation Ð methods: statistical Ð radio lines: general
\end{keywords}

\section{Introduction}
\label{sec:intro}

During the Epoch of Reionization (EoR), gas in the Universe reionized
after having been neutral for about 500 Myr during the
so called â Dark Ages. The EoR
is thought to be caused by the first radiating sources, and its study
is crucial to our understanding of the physics of these sources and
how they influenced the formation of later generations of
astrophysical objects. 
Current observational constraints indicate that the EoR occurred at $6.5 \lsim
z \lsim 12$, as inferred from SDSS high redshift quasar spectra \citep{fan03,fan06}, WMAP \citep{page07}, SPT \citep{zahn12} and IGM temperature
measurements \citep{theuns02, bolton10}. 
In addition, recent HST observations of the Hubble Ultra Deep Field taken with the new 
Wide Field Camera 3 (WFC3) found a large sample of Lyman break galaxies at $7\lsim z \lsim10$ \citep[see e.g., ][]{oesch10, bouwens10, bunker10}.
These authors found that these galaxies do not produce enough ionizing photons to account for the
Universe's full reionization by redshift 6 and concluded that an additional source of ionizing photons is 
required \citep{bouwens11}. Furthermore, measurements of the number of ionizing photons per baryon from Lyman $\alpha$
forest spectra at $z\approx 6$ yields a low number of such photons ($\sim 3$), that is,
 the reionization process is photon starved \citep{bolton07, calverley11}. When combined, these measurements lend themselves to the
notion that reionization is a slow and  drawn out process.

Our current best hope to
study this epoch in detail lies in observations of the redshifted HI
21cm emission line \citep[see e.g., ][]{madau97, shaver99, furlanetto06a, pritchard12}. 
To date, a number of experiments are planning to measure the EoR
with the redshifted 21 cm line (e.g. LOFAR$^1$, GMRT$^2$, MWA$^3$, 21CMA$^4$, PAPER$^5$).
These experiments seek statistical detection of the
cosmological 21cm signal, with the most widely studied such statistics
being the rms and the power spectrum of the brightness temperature
and their evolution with time (e.g. \citet{morales04, barkana05,
 mcquinn06, bowman06,  pritchard07, jelic08, harker09b, harker10, pritchard08}). Though recently, \citet{datta12} have shown that 
 one can image large bubbles of ionization around very high redshift powerful quasars with LOFAR.
 In particular, \citet{jelic08}, \citet{harker10} and more recently \citet{chapman12} showed that despite the low signal-to-noise ratio,
prominent foregrounds and instrumental response, the 21~cm
rms and power spectrum can be extracted from the data collected with the Low
Frequency Array (LOFAR). Similar studies have been carried out for the MWA case \citep{geil08, geil11}.
\footnotetext[1]{Low
Frequency Array, http://www.lofar.org/}\footnotetext[2]{Giant Metrewave Telescope,
http://www.gmrt.ncra.tifr.res.in/}\footnotetext[3]{Murchison Widefield
Array, http://www.haystack.mit.edu/ast/arrays/mwa/}\footnotetext[4]{21 Centimeter
Array, http://web.phys.cmu.edu/\~{}past/}\footnotetext[5]{Precision
Array to Probe the EoR,
http://astro.berkeley.edu/\~{}dbacker/eor/}

The current generation of telescopes are designed to detect the EoR statistically, rather than image it, 
for a number of reasons. On small scales the noise level per resolution 
element is relatively high. For example, at 150~MHz LOFAR  will have a 56~mK system noise per resolution
element ($\approx 3$ arcmin) after 600 hours of observations with a 1~MHz bandwidth, corresponding to a signal-to-noise
ratio of $\sim 0.2$ at these scales.
 On large scales there are two issues. The first one has to do with the
typical sizes of ionized and neutral regions at each redshift, which limits the maximum scale at which
smoothing of the data remains useful. It can also be shown that during the dark ages (no ionization bubbles) 
smoothing on very large scales does not help so much due to the low level of contrast and to the fact that beyond 1 degree the power due to cosmological fluctuations drops
very quickly \citep[see e.g.,][]{santos05, jelic08}.
The other, and potentially more severe,  issue is that of  the foregrounds which  dominate the measurement 
on all scales and  become even more prominent on large scales with power increasing as $\theta^{2.5-3}$ \citep{tegmark00,
giardino02, santos05, jelic08, jelic10b, bernardi09, bernardi10}. This means that on large scales the influence of these foregrounds will be harder 
to filter out. The common wisdom is that scales beyond half a degree will remain inaccessible even after a number of years of observations.

In this paper we argue that, despite the hurdles we have listed, imaging of the EoR
on very large scales is possible with the current generation of telescopes. In presenting our case we rely 
on two arguments.
Firstly, inspection of large scale EoR simulations ($\gsim 200\cMpch$)  shows that towards the end of the reionization process
one can still find sufficiently large neutral patches so as to allow
imaging of this process after smoothing on sufficiently large scales. For example, a neutral patch 
towards the end of reionization with a scale of roughly 120 comoving $\cMpch$ ($1^\circ$) would have a signal-to-noise ratio per
smoothing cell, after 600 hours
of observations,  of about 4 since it would have about $20\times 20$ independent resolution elements ($1^\circ/3'=20$).
Notice that this argument would not work early in the reionization process since the variation in the 21~cm
intensity in a given field, measured by radio interferometers, will be driven by the cosmological density fluctuation field, $\delta$,
which is relatively small -- unless the spin temperature itself exhibits fluctuations above and below the CMB
temperature \citep{pritchard07, pritchard08, pritchard10, baek10, thomas11}.  Towards the later stages of the reionization process, however, the variations in 
the intensity are driven by the difference between neutral and ionized regions.  The 120 $\cMpch$ scale is driven 
mainly by the clustering scale of the Universe's large-scale structure, which the ionization sources, independent of their nature, tend to follow.

Secondly, the current state-of-the-art foreground fitting methods, such as Wp smoothing \citep{harker10} or Independent Component Analysis 
\citep{chapman12},  do a very good job even on large scales,
rendering them accessible for EoR analysis. The availability of such techniques together with the existence of the large-scale neutral patches towards the end of reionization, make it
possible to image the EoR from 21~cm data on large scales.

This paper is organized as follows. In section~\ref{sec:signal} we describe the cosmological signal and its basic equation. In section~\ref{sec:simulations}  we introduce the very large-scale simulations needed to demonstrate our argument. In these simulations we also include the influence of the telescope response, noise and foregrounds.  In section~\ref{sec:results} it is shown that, based on these large-scale simulations, imaging of the EoR on large scales is indeed possible with current instruments. 
We also argue that the key to successful imaging on large scales is the ability to remove the foreground
signal with sufficient accuracy on scales larger than 1 degree.  The paper concludes with a summary and outlook (\S~\ref{sec:conclusions}).

\section{Cosmological 21~cm signal}
\label{sec:signal}
In radio astronomy, where the Rayleigh-Jeans law is usually
applicable, the radiation intensity, $I (\nu)$, is expressed in terms
of the brightness temperature, so that
\begin{equation} I (\nu) = \frac{2 \nu^2}{c^2} k_{B} T_b,
\end{equation} where $\nu$ is the radiation frequency, $c$ is the
speed of light and $k_{B}$ is Boltzmann's constant \citep{rybicki86}.
This in turn can only be detected differentially as a deviation from the
Cosmic Microwave Background (CMB) temperature, $T_{CMB}$. The predicted
differential brightness temperature $\delta T_b \equiv T_b -T_{CMB}$, which reflects the
fact that the only meaningful brightness temperature measurement, insofar as the intergalactic medium (IGM) is concerned, is when it deviates from $T_{CMB}$. 
Derivation of $\delta T_b$  yields  \citep{field58, field59b, madau97, ciardi03a},
\begin{eqnarray} \delta T_b &  \approx    28 \mathrm{{mK}} &  \left( 1 + \delta
\right)  x_{\small{HI}} \left( 1 - \frac{T_{CMB}}{T_{spin}} \right) \left(\frac{dv_r/dr}{H(z)}+1 \right)^{-1}
 \nonumber\\ & &  \times \left( \frac{\Omega_b h^2}{0.023} \right)
\sqrt{\left( \frac{1 + z}{10}\right)
\left(\frac{0.15}{\Omega_m h^2}\right)
}, 
\label{eq:dTb}
\end{eqnarray} where $h$ is the Hubble constant in units of $100~
\mathrm{{km} \, s^{- 1} {Mpc}^{- 1}}$, $\delta$ is the mass density
contrast, $v_r$ is the line-of-sight velocity component, $x_{HI}$ is the neutral fraction, $\Omega_m$ and
$\Omega_b$ are the mass and baryon densities in units of the critical
density and $H(z)$ is the Hubble parameter. Note that the three quantities, $\delta$, $v_r$, $x_{HI}$ and
$T_{s}$, are all functions of 3D position.

Equation~\ref{eq:dTb} shows that the differential brightness temperature is composed of a mixture of cosmology dependent 
and astrophysics dependent terms. The equation is clearly complex yet at the same time
information rich. This is simply because at different stages in the evolution of reionization
$\delta T_b$ is dominated by different contributions. For example, at certain redshift ranges 
no significant ionization has taken place, i.e. $x_\MHI \approx 1$ everywhere,  yet there is enough heating to
render $T_{spin} \gg T_{CMB}$. In such case, the brightness temperature 
is proportional to the density fluctuations making its measurement an excellent probe of cosmology.
However, at low redshifts ($z\lsim 9$) a significant fraction of the Universe is expected to be ionized and the measurement
is dominated by the contrast between the neutral and ionized regions, hence, probing the 
astrophysical source of ionization (see e.g., \cite{iliev08, thomas09, thomas11}). During the stage at 
which reionization occurs it is safe to assume that $T_{spin}\gg T_{CMB}$~\citep{pritchard08, pritchard10}.

\begin{figure*}
\centering
\includegraphics[width=1.0\textwidth]{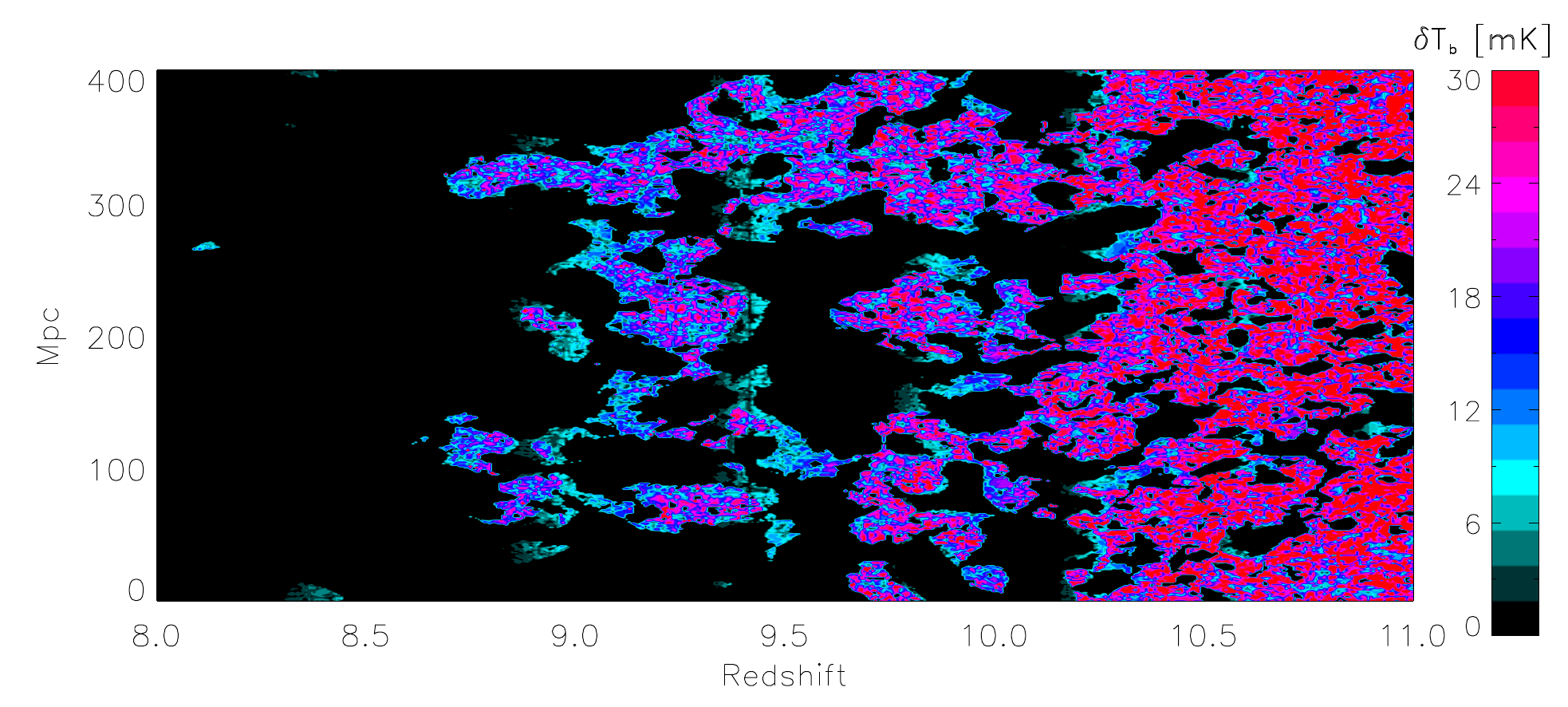}
\caption{{A slice of $\delta T_b$ along the frequency direction
for the 400~$\cMpch$ simulation obtained with  {\tt\small 21cmFAST}. The redshift at which the
box is half ionized is 10. We will show results from redshift 9 in this paper where the neutral 
fraction is 0.2. The color scale is shown to the right of the figure.}
}
\label{fig:slice}
\end{figure*}

Since interferometers do not measure the mean $\delta T_b$ but rather are sensitive to its fluctuations, it
is easy to see that early in the reionization process, when the ionized regions are quite negligible, the rms 
of the measurement will be driven by $\delta$,  the cosmological density contrast fluctuations (assuming $T_{spin} \gg T_{CMB}$).
Later in the reionization process, however, when the typical size of the ionized regions becomes larger than
the interferometer's resolution scale, the measured signal is driven by $x_{\HI}$, which increases the contrast and
as a result the ability to observe the signal. 

In this paper we ignore the issue of calibration errors.
This issue might become important for the real data \citep{datta10}, but so far the indication is that the calibration will be achieved 
with very  high accuracy \citep[][]{yatawatta09, kazemi11} as shown recently by observation of two LOFAR fields
(Yatawatta and Labropoulos, private communication). 

\section{Data Simulations}
\label{sec:simulations}

Since here we argue that instruments such as LOFAR will enable imaging the EoR on large scales,
it is important to test the influence of the noise, instrument response, foreground extraction, and the 
distribution of the 21 cm signal power on various scales as a function of redshift.
Hence one needs to explore scales well in excess of $\approx 120 \cMpch$, namely, about 1 degree on the sky. This is 
the natural scale of the large-scale structure and is marked in the cosmological power spectrum by the turn over from
$P(k) \propto k^n$ to $P(k) \propto k^{n-4}$, where the primordial power law index $n\approx 1$. This scale is equal to
the comoving horizon size at the era of equality between matter and radiation.
  
Therefore, as a first step, we have to create very large-scale cosmological reionization simulations, from which a $\delta T_b$
signal cube is produced (signal as a function of frequency). We then add to this cube a realistic model of foregrounds, 
expected instrumental response and (system) noise.

\subsection{Large-scale 21~cm simulations}
Full radiative transfer simulations on very large scales -- in excess of 200 $\cMpch$ -- are not available. Hence,  recourse must be had to semi-analytical methods.
Here we make use of two sets of simulations. The first simulations are produced using the publicly available package {\tt\small 21cmFAST} \citep{mesinger10, mesinger07, zahn07} which uses a semi-analytical 
approach to produce very large-scale simulations of the reionization process. See also \citet{santos10} who uses an alternative method 
to create fast large-scale reionization and redshifted 21 cm simulations, called {\tt\small SimFast21}.

{\tt\small 21cmFAST} uses perturbation theory, the excursion set formalism,
and analytic prescriptions to generate evolved 3D realizations of the density, ionization, peculiar velocity and spin temperature fields, which it then combines 
to compute the 21-cm brightness temperature. The method has been thoroughly tested against more accurate reionization
codes \citep{mesinger10}.

We produce three-dimensional 400~$\cMpch$ simulation boxes of the brightness temperature from redshift 12 down to redshift 6. The 
resolution of the simulation box is 1~$\cMpch$.  The simulations assume the standard WMAP cosmological parameters \citep{spergel07}. The simulation box is binned with a $400^3$ grid. The output of the code includes
the spin temperature, the ionization fraction, the kinetic temperature and peculiar velocity field. These outputs are used
by {\tt\small 21cmFAST}  to create a brighness temperature box. For more details
please see \citet{mesinger10} and \citet{mesinger07}. We then use the method developed by \citet{thomas09}
to create an observational box of the 21~cm signal spanning the frequency range of $115-200~\mathrm{MHz}$. 
A slice through the signal cube along the frequency direction is shown in Fig.~\ref{fig:slice}.
In this simulation the ionization process reaches its mid point at $z\approx 10$. We will present a number of plots in the
remainder of the paper at $z=9$ where the neutral fraction is 0.2.

The second set of simulations we use to test the imaging possibility is based on the EoR simulations program
called {\tt\small\ BEARS} \citep{thomas08, thomas09}. This scheme includes the physics of reionization
in more detail relative to {\tt\small 21cmFAST} but needs cosmological simulations in order to create the reionization history, which makes
it more time consuming than  {\tt\small 21cmFAST}. {\tt\small\ BEARS} assumes a spherically symmetric ionization region around
each ionizing source but can easily allow for a wide range of sources with very different spectral energy distributions (SEDs). 
Here we use a dark matter simulation of $512^3$ particles in a cube with comoving side length of  $200\cMpch$ with WMAP standard parameters. The sides thus have twice the length of the
simulations shown in \citet{thomas09} and used in our previous work on
LOFAR EoR signal extraction \citep{harker09b, harker09a}.
This leads to a minimum resolved halo mass of around $3\times 10^{10}\ h^{-1}\
\mathrm{M}_\odot$. Dark matter haloes are populated with sources whose
properties depend on some assumed model. For this paper we explore both the
`quasar-type' and the stellar source models of \citet{thomas08, thomas09, zaroubi07}.
The topology and morphology of reionization is different in the two
source models. We might expect quasar reionization to allow an
easier detection than stellar reionization, since the regions where the
sources are found are larger and more highly clustered, producing
larger fluctuations in the signal. This is used here as a check on whether a completely 
different and more detailed simulation scheme yields similar conclusions to the ones based 
on {\tt\small 21cmFAST} simulations. 

Unfortunately, a full and detailed radiative transfer simulation on such a scale is not publicly available
at this stage and we expect our conclusions to vary somewhat with higher resolution simulations.
Still, the main results are expected to remain valid, as is indeed seen in such a simulation that is 
currently being analyzed by Iliev \& Mellema (private communication).

\subsection{Instrumental response}

In radio interferometry the measured spatial correlation of the electric field between two
interferometric elements (stations) $i$ and $k$ at a given time, $t$, is called the visibility and is given
by \citep{taylor99, thompson01}:
\begin{equation}\label{vis}
  V^{i,k}_{\nu}({u,v;t})=\int I^{i,k}_\nu(l,m;t)e^{2\pi j({u}l+{v}m)}dldm,
  \label{eq:visibility}
\end{equation}
where $I^{i,k}_\nu$ is the  observed intensity at frequency $\nu$ observed by correlating stations $i$ and $k$ and $j=\sqrt{-1}$.
The coordinates $l$ and $m$ are the projections (direction cosines) of the source in terms of  the baseline\footnote{We ignore the effect
of the Earth's curvature, the so called w-projection.}. The size of the station gives the resolution, i.e., minimum uv cell size,
at which the uv-plane is covered.
From this equation it is evident that the observed visibility
is basically the Fourier transform of the intensity measured at the coordinates u and v (uv-plane).
Following \citet{harker10}, we define a sampling function,
\begin{equation}
S_\nu(u, v)\equiv \sum_{\forall k\in \mathrm{pixel}\, \left(u,v\right)} \int \delta^{D}(u'-u_{k},v'-v_{k};t, {\nu})\mathrm{d}t \mathrm{d}u'\mathrm{d}v'
\end{equation}
 which gives how a distribution of  interferometer baselines 
sample Fourier space during the time of observation. Here $\delta^D$ is the Dirac delta function, $k$ is a uv-track of a baseline.

Obviously,  Eq.~\ref{eq:visibility}  indicates that the sampling function depends on frequency as well as on the number
of baselines and their distribution, i.e., how the visibilities are distributed in the uv-plane.
Here we  assume that the uv coverage is the same at all frequencies. In practice, a uniform uv coverage at all
frequencies could be achieved by ignoring all the the uv points that are only partially covered  within
the frequency range of interest.  This would require discarding some of the data - in the case of 
LOFAR one loses approximately 20 per cent of the data. This has two effects: an increase the level of noise and
a reduction in the resolution at high frequencies. However, the assumption here is that the uv coverage of the 
telescope is quite dense  and complete. In practice, all the baselines data, including from the long ones,
 will be used to model and remove compact sources.

In the case of LOFAR, to simulate our data in the uv plane we perform a two-dimensional
Fourier transform on the image of the foregrounds and
signal at each frequency, and multiply by a mask (the uv coverage)
which is unity at grid points in Fourier space (uv cells) where
$S(u, v) > 0$, and is zero elsewhere.

\begin{figure}
\centering
\includegraphics[width=0.5\textwidth]{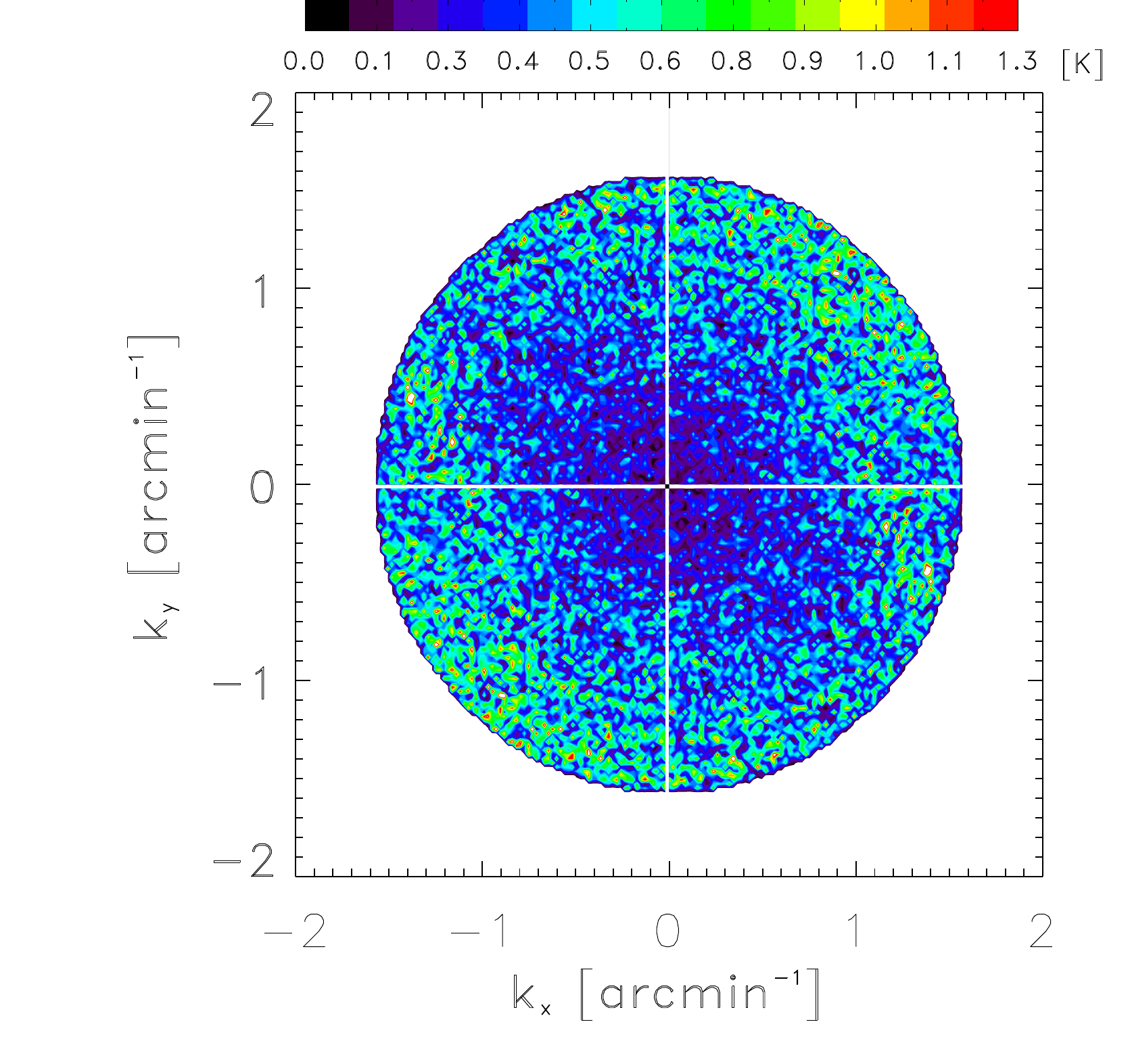}
\caption{{A noise realization in the uv-plane for 600 hours of observation.
The absolute value of the noise is drawn here and shown in  K (and not mK). The rms of the noise in the image plane is about 56~mK.
In order to convert from Jy to mK we assume that $k=2\pi/\theta$. 
The truncation in the map is made at a radius corresponding to 4 arcmin.}
}
\label{fig:noisemap}
\end{figure}

\subsection{Noise}

Assuming the noise in the measurement follows a Gaussian distribution for each component, the 
uncertainty in the measurement of the visibility at a given uv-plane pixel is inverse proportional to 
$\sqrt{S}$.
The noise realization in the uv-plane is created at each uv point by drawing from a 
complex Gaussian field with an rms proportional to 
$1/\sqrt{S}$ for all the cells within the mask down to a uv distance
that is equivalent to 4 arcmin resolution. The realization in the uv-plane is done so as to fulfill the reality condition.  
At scales smaller than this we truncate the sampling.
The truncation is performed in order to avoid the noise normalization being controlled by
the very low number of samplings at the edge of our uv sampling, namely, at the limit
in which the Poisson-noise can not be approximated by white Gaussian noise.
The noise realization in the image plane can then be obtained by inverse Fourier transform 
the uv-plane noise.
The overall normalization of the level of noise is chosen
so that the noise images have an rms in the image-plane  of 56 mK on an image using
1~MHz bandwidth at 150~MHz at the resolution limit. This is the rms expected from LOFAR after
 600 hours of observation of one EoR window with one synthesized beam. 
 The noise level depends on the system temperature which is assumed
to be $T_{sys} = 140 + 60\times (\nu/300~\mathrm{MHz})^{-2.55}$ K~\citep{jelic08}.
A much more detailed account of the calculation of noise levels
and the effects of instrumental corruption for the LOFAR EoR
project may be found in \citet{panos09}.

Fig.~\ref{fig:noisemap} shows a noise realization used in the simulations. The circle beyond
which there is a cut in the uv data corresponds to 4 arcmin resolution at 150~MHz. The distribution
of the LOFAR stations was chosen to ensure a constant noise at each scale for a measurement
of the power spectrum. This is why the noise level per pixel increases as a function of uv radius. Remember
that the experiment was optimized for a power spectrum measurement and not for imaging.

\begin{figure*}
\centering
\includegraphics[width=0.49\textwidth]{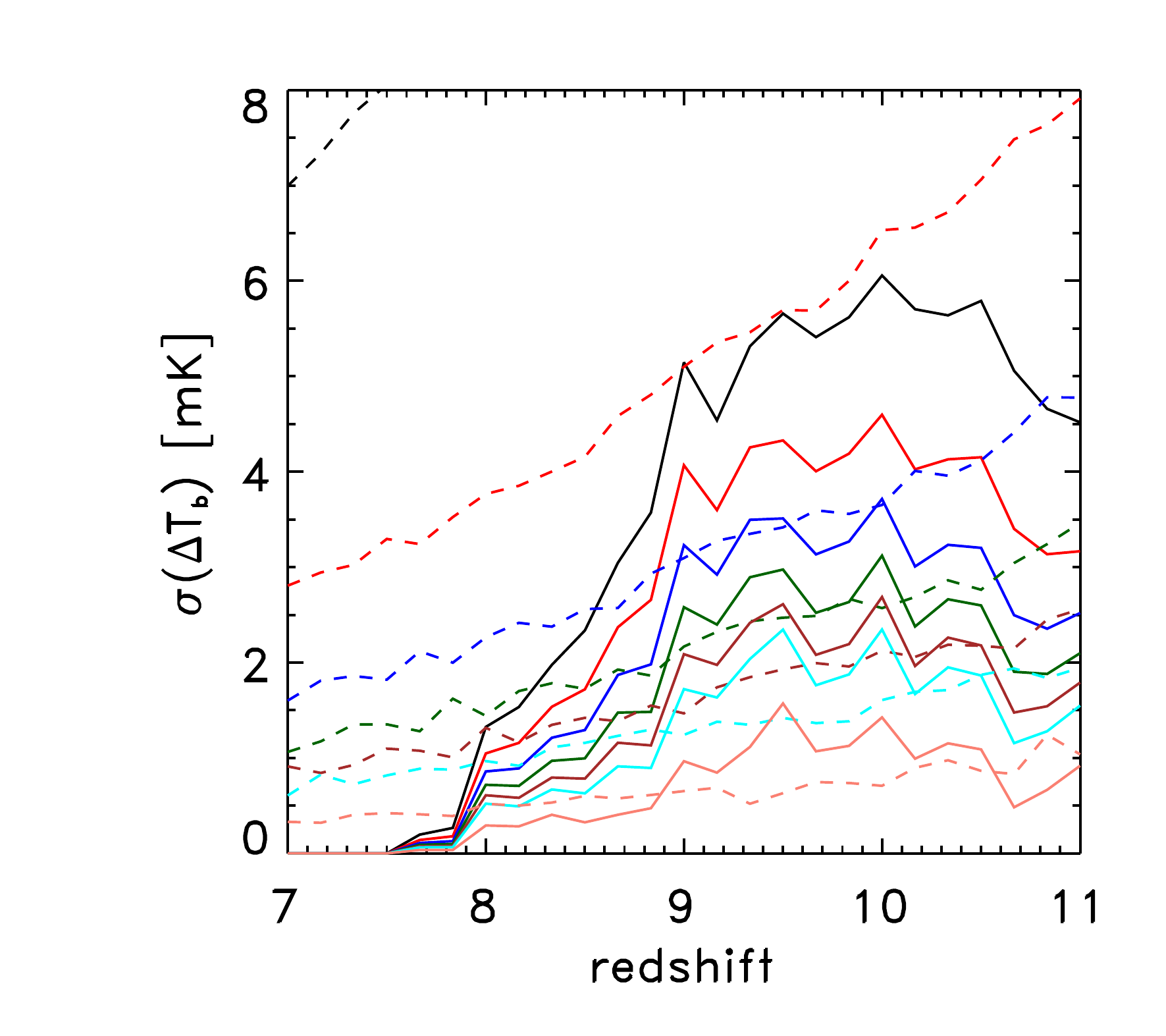}
\includegraphics[width=0.49\textwidth]{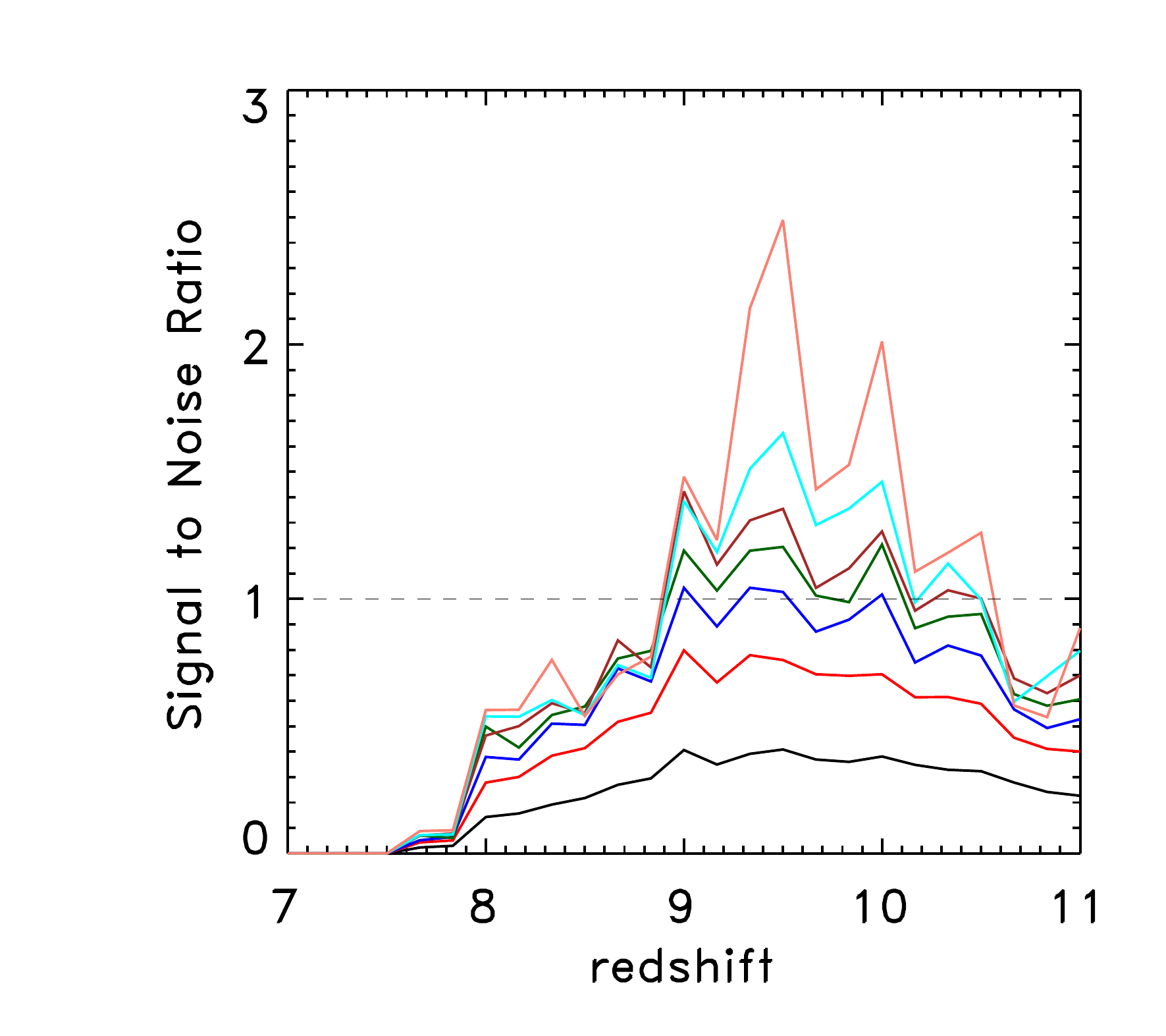}
\caption{{The left panel shows the standard deviation of the signal (solid lines) and the noise (dashed lines) as a function of redshift for the 
400~$\cMpch$ simulation noise and signal. Here we assume 600 hours of observation. The black, red, blue, dark green, brown, cyan and pink solid lines are for instrument resolution ($\approx 3$ arcmin.), 5, 10, 15, 20, 25 and 30  arcmin resolutions, respectively. 
The right panel shows the signal-to-noise ratio for the same cases.
Notice that with 15 arcmin resolution and above there is a redshift range in which the signal rms exceeds the noise rms. This high signal-to-noise region is typically centered around the redshift at
which the IGM is 50\% ionized.
It is also worth noting that although the rms of the signal decreases with the smoothing scale, the decrease in the
noise is even larger.}
}
\label{fig:rmsz}
\end{figure*}

\subsection{Foregrounds and extraction}
As mentioned earlier, a very important ingredient to consider here
is the accuracy of extracting the foregrounds, especially on large 
scales. This is because the foreground's power increases with scale and
there is no guarantee that the extraction algorithm, which fits the foregrounds
along the frequency direction at small spatial scales, will not leave any
large-scale residuals. 

For the foregrounds we use the simulations of \citet{jelic08, jelic10b}. These
incorporate contributions from Galactic diffuse synchrotron and
free-free emission, and supernova remnants. They also include 
unresolved extragalactic foregrounds from radio galaxies and radio
clusters. We assume, however, that point sources bright enough to
be distinguished from the background, either within the field of
view or outside it, have been removed well below the noise level from the data. Observations
of foregrounds at 150~MHz at low latitude \citep{bernardi09, bernardi10} 
indicate that these simulations describe the properties
of the diffuse foregrounds well.

To test the foregrounds' influence, especially with the amount of noise in the data, we apply an extraction
algorithm on mock data. The mock data include the simulated cosmological signal,
the instrument response, noise and foregrounds. As mentioned earlier,
 calibration errors and other systematics are not taken into account in this
simulation.

In order to extract  the foregrounds, we assume that they have no small-scale features
along the frequency direction. Given what we know about the physical origin of the 
galactic and extragalactic foregrounds, this assumption is quite reasonable \citep{petrovic10}.
For the extraction we use the W$_p$ algorithm which is a non-parametric
method that is very suitable for fitting the spectrally smooth foregrounds in
EoR data sets. The method was developed for general cases by \citet{machler95},
 and has been used by \citet{harker09a} as an algorithm for 
fitting EoR foregrounds. Briefly, the method is a penalized maximum likelihood algorithm
that is designed to find the maximum likelihood fit for the data but penalizes relative change of curvature.
That is so to say, the method finds the best fit curve to the data with minimum the smallest possible 
ruggedness.

The W$_p$ method has shown very good results for fitting the foregrounds
both in real space (image-plane) and in Fourier space (uv plane) for up to 100 $\cMpch$
and its influence on the power spectrum statistic has been tested and shown not to be significant
up to the scales of the simulation \citep{harker10}. Here however, we would like to test it on
much larger scales  than considered previously. In this paper, application of this method is done using uv-plane fitting
which has shown slightly better results than image-plane fitting  \citep{harker10}. 

One should note that using predetermined functions, e.g., polynomials,
 to fit the data might introduce systematics due to over- or under-fitting of the foregrounds. 
 Hence, the use of more advanced non-parametric techniques is essential in this case \citep[see e.g., ][]{harker09b, chapman12}.

\section{Results}
\label{sec:results}

\begin{figure}
\centering
\hspace{0cm}
\includegraphics[width=0.51\textwidth]{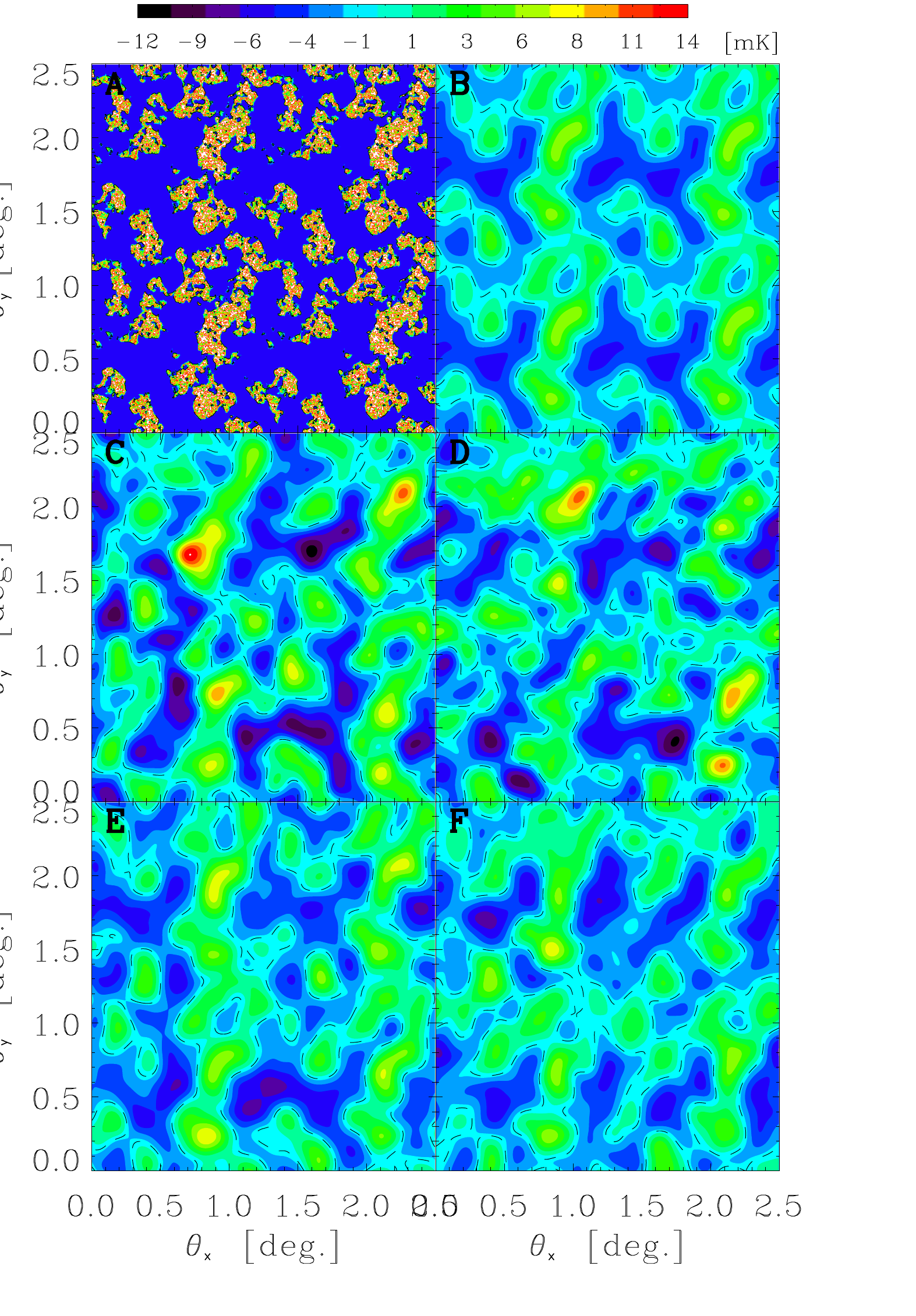}
\vspace{.2cm}
\caption{{EoR maps at redshift 9, with a $\langle x_{HI} \rangle=0.2$ field of view of  $2.5^\circ\times2.5^\circ$ (note that the LOFAR
field of view will be around $5^\circ\times5^\circ$).
 Panel A shows the original simulated EoR map at full resolution. The map shown in Panel B
is smoothed with a 20 arcmin (standard deviation) Gaussian kernel. Panel C shows the same map as in B but with noise added to it 
assuming 600 hours of observation with  LOFAR. Panel D is 
 the same as C, i.e., with 600 hours of observation noise, but here the foregrounds were added and then extracted with the W$_p$ fitting procedure \citep{harker09a}. Panel E is the same as C but with half the noise level of the map in panel B (2400 hours of observation). Panel F is the same but with 2400 hours of observation noise and foregrounds that were added and then
 extracted with the W$_p$ fitting procedure.
 The contour levels are colour coded as shown in the colour table at the top of the figure.
}}
\label{fig:maps}
\end{figure}

We plot, in Fig.~\ref{fig:rmsz}, the standard deviation of the smoothed 
cosmological signal (solid lines) and the smoothed noise field (dashed lines) as a function of redshift. The smoothing is done with a Gaussian 
kernel of 5, 10, 15, 20 and 25 arcmin. It is clear that for smoothing scales $\gsim 15$ arcmin there is a redshift range in which the
signal becomes larger or comparable to the noise. Obviously, if the cosmological signal is coherent on scales larger than
the smoothing scales, these structures will show up as EoR features in the 21~cm maps. 

The fact that the EoR signal catches up with the noise at large smoothing scales is driven by the
the existence of very large ionized and neutral patches. 
Regardless of their nature, the sources that drive the ionization bubbles follow
the large-scale structure which has a natural scale of $120~\cMpch$
This sets the largest scale up to which one can still see coherent structures of neutral and 
ionized regions.

Since 120~$\cMpch$ is the natural scale of the large-scale structure, one can ask why such smoothing 
does not facilitate imaging of the EoR at all redshifts? The answer is simply that without the ionized regions
the contrast within the map is driven by the underlying density. This of course assumes that one can ignore the 
fluctuations due to $T_{spin}$, which a good assumption in the redshift range of 6-11.5 that LOFAR will probe, but
not a good assumption at much higher redshifts \cite{}. That is to say, if the brightness temperature 
fluctuations were driven by the cosmological density alone, then it would be more difficult, with LOFAR, to image the signal even on large scales.

Fig.~\ref{fig:maps} shows original and reconstructed maps of the simulated EoR signal in a $2.5^\circ\times2.5^\circ$ field of view
at redshift 9.  We note here that this is about a quarter of the LOFAR field of view assuming a single beam.
We compare here the 20 arcmin Gaussian smoothed map shown in Panel B with the following cases: 
Panel C, with a noisy signal assuming 600 hours of integration with LOFAR but without including
foreground effects. Panel D, noisy signal assuming 2400 hours of integration of the same field (half the noise level), still
without the inclusion of the foreground effects. Panel E, the same map as in C, i.e., with noise added assuming 600 hours of integration,
 but with inclusion of the foregrounds and their extraction with the W$_p$ fitting procedure \citep{harker09b}. 
The smoothing is done with a 20 arcmin Gaussian kernel. After smoothing the signal rms in map A is 2.6~mK. The noise level 
 after 600 hours of integration in maps C and D is is 2.~mK and 2.2~mK (0.9~mK of which are due foregrounds residual), respectively.
The noise levels after 2400 hours of integration and 20~arcmin smoothing in maps E and F are 1.1~mK and 1.3~mK (0.7~mK of which due to 
foreground residuals), respectively.
It is clear that after 600 hours of 
observation, one has the ability, albeit a limited one, to map the EoR signal, and the contour map is dominated by the noise. However,
Panel F shows that after 2400 hours of observation the noise influence drops significantly at this smoothing scale, and 
 a more reliable map can be seen. This remains true even after inclusion of the foreground effect, that is, when traces of the foreground extraction are present on large scales.
 
A visual inspection shows a clear similarity between Panel B and Panels C-F in Fig.~\ref{fig:maps}. To
quantify this similarity we use two different methods. The first method is the Pearson cross-correlation coefficient,
$\rho$, calculated with the formula,
\begin{equation}
\rho=\frac{\sum_{i}(x_i - \left<x\right>)(y_i-\left<y\right>)}{\sqrt{\sum_i(x_i - \left<x\right>)^2}\sqrt{\sum_i(y_i-\left<y\right>)^2}},
\label{eq:spearman}
\end{equation}
where $x_i$ and $y_i$ are  the value of the pixel, $i$, in the two maps. 
The results of this calculation are shown in Table~\ref{table:spearman}.
The table shows the correlation coefficient between the maps C-F and
the noise- and foreground-free map B from Fig.~\ref{fig:maps}. The existence of 
foregrounds and their extraction in maps D and F clearly reduces the correlation. Also the
higher noise in maps C and D results in smaller correlation coefficient. Still, the correlation
coefficients  shown in the table are very high in all cases.

\begin{table}
\caption{Spearman Correlation Coefficients between Map B and the other maps.}
\begin{tabular}{c  c  c  c c}
\hline\hline\\
Map & C &  D & E & F\\
$\rho$ & 0.77 & 0.68 & 0.93& 0.80\\
\hline
\end{tabular}
\label{table:spearman}
\end{table}

\begin{figure}
\centering
\hspace{0cm}
\includegraphics[width=0.45\textwidth]{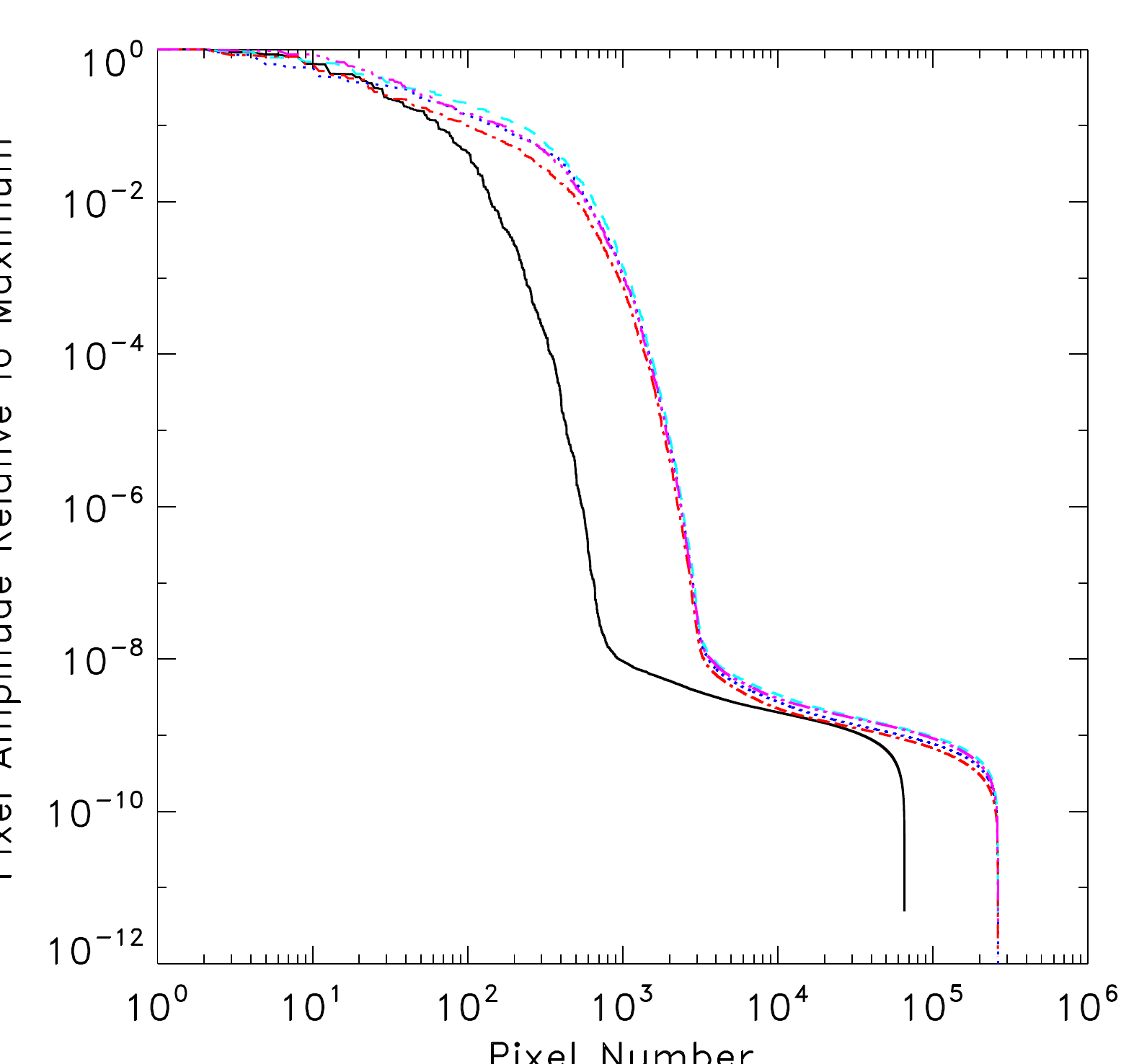}
\vspace{.2cm}
\caption{{
The rank-ordered Fourier space amplitude for each of the 5 images,
B-F, shown in Fig.~\ref{fig:maps}. All curves are normalized with respect to their maximum amplitudes.
The solid black line is plot for map B whereas the other curves show the Fourier space 
amplitudes of image C (blue dotted line), D (cyan dashed line), E (red dotted-dashed line) and F (magenta double dotted-dashed line). All $uv$-maps are dominated by the highest few hundred pixels. The
rest of the Fourier space pixels are noise dominated as demonstrated by the sudden drop in the amplitudes and their
almost flat slope thereafter. }}
\label{fig:amplitudes}
\end{figure}

The other method we use to quantify the correlation between the maps is to inspect
their phase information. This is done by Fourier transforming each map and then checking
whether the Fourier space phases of the maps C-F correspond to those of the original map B.
If this were done to all the points in Fourier space, then one would obtain no correlation between the phases.
This is because the amplitudes of most points in the Fourier transform of maps C-F is dominated
by numerical noise and contain no useful information. In 
Fig.~\ref{fig:amplitudes} we show  a log-log plot of the rank-ordered Fourier coefficient amplitudes of the five images.
The solid black line is the one for the image shown in Panel B of Fig.~\ref{fig:maps}, whereas the others
are for the rest of the 20 arcmin smoothed images. Each line is normalized such that its maximum 
amplitude is one. All the lines show the same typical behavior where the amplitude of the first few hundred pixels is high 
but then it drops exponentially to slowly varying values (almost flat) which is typical white noise
behavior.  Note that the number of significant coefficients is larger in  maps C-F than in map B 
because the former contain a contribution from the (correlated) system noise. The flatness of this part of 
the plots indicates that they are dominated by white noise. We would like to emphasize that this is not the system noise contribution which typically has a much larger amplitude and is not white (see Fig.~\ref{fig:noisemap}). 
Therefore, in order to compare the phases of the 
various images, we only take into account the pixels that have values larger than $10^{-4}$ times the maximum amplitude.

\begin{figure}
\centering
\hspace{0cm}
\includegraphics[width=0.49\textwidth]{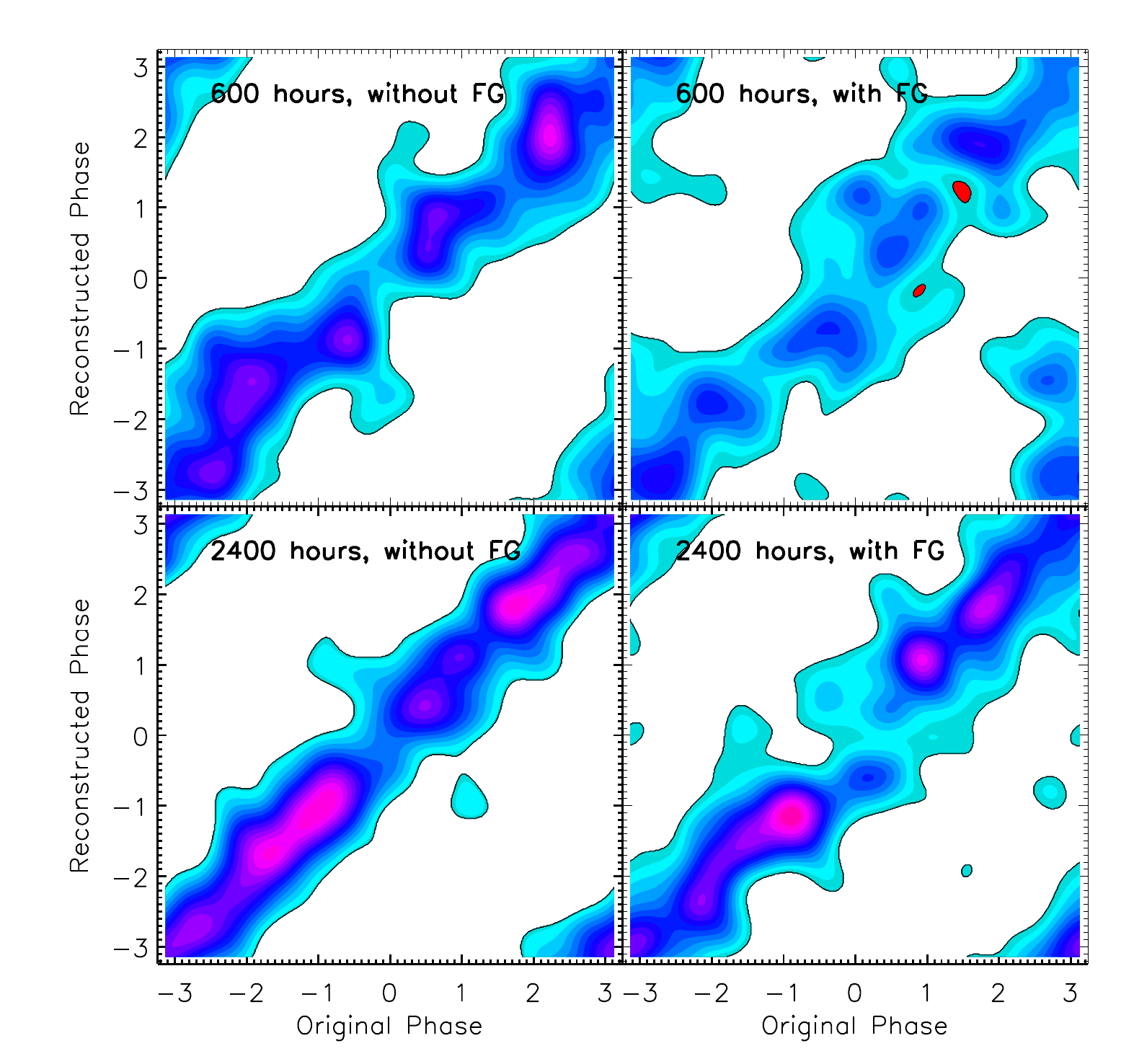}
\vspace{.2cm}
\caption{{
The phases of the reconstructed images (maps C-F in Figure~\ref{fig:maps})
versus the phases of the original map (map B in Figure~\ref{fig:maps}). The plots are shown as density
plots where the density represents the number of point per unit area, hence, highly correlated maps should show 
as high density contours at  the diagonal. The map shows the highest 67\% of the density PDF. The high correlations at the upper left side and lower right side 
of the figures simply reflect the periodicity of the phases. }}
\label{fig:phases}
\end{figure}

Next, we plot the phases of the pixels with relatively high amplitudes ($\geq 10^{-4}$ of the maximum amplitude).
Each of the four panels of Fig.~\ref{fig:phases} plots the phases of the reconstructed images (maps C-F in 
Fig.~\ref{fig:maps})
versus the phases of the original map (map B in Fig.~\ref{fig:maps}). These plots are presented as density 
plots. A high correlation shows as high concentration of points (high contour values) along the diagonal. In all 
the panels the correlation between the phases is obvious. The best correlation is clearly obtained in the
lower left panel because map E has the lowest noise and is without foregrounds. The worst correlation, though
still a very clear correlation, is obtained in the upper right panel because map D has high noise and still has 
some residuals from the subtraction of the foregrounds. 

\begin{figure}
\centering
\hspace{0cm}
\includegraphics[width=0.52\textwidth]{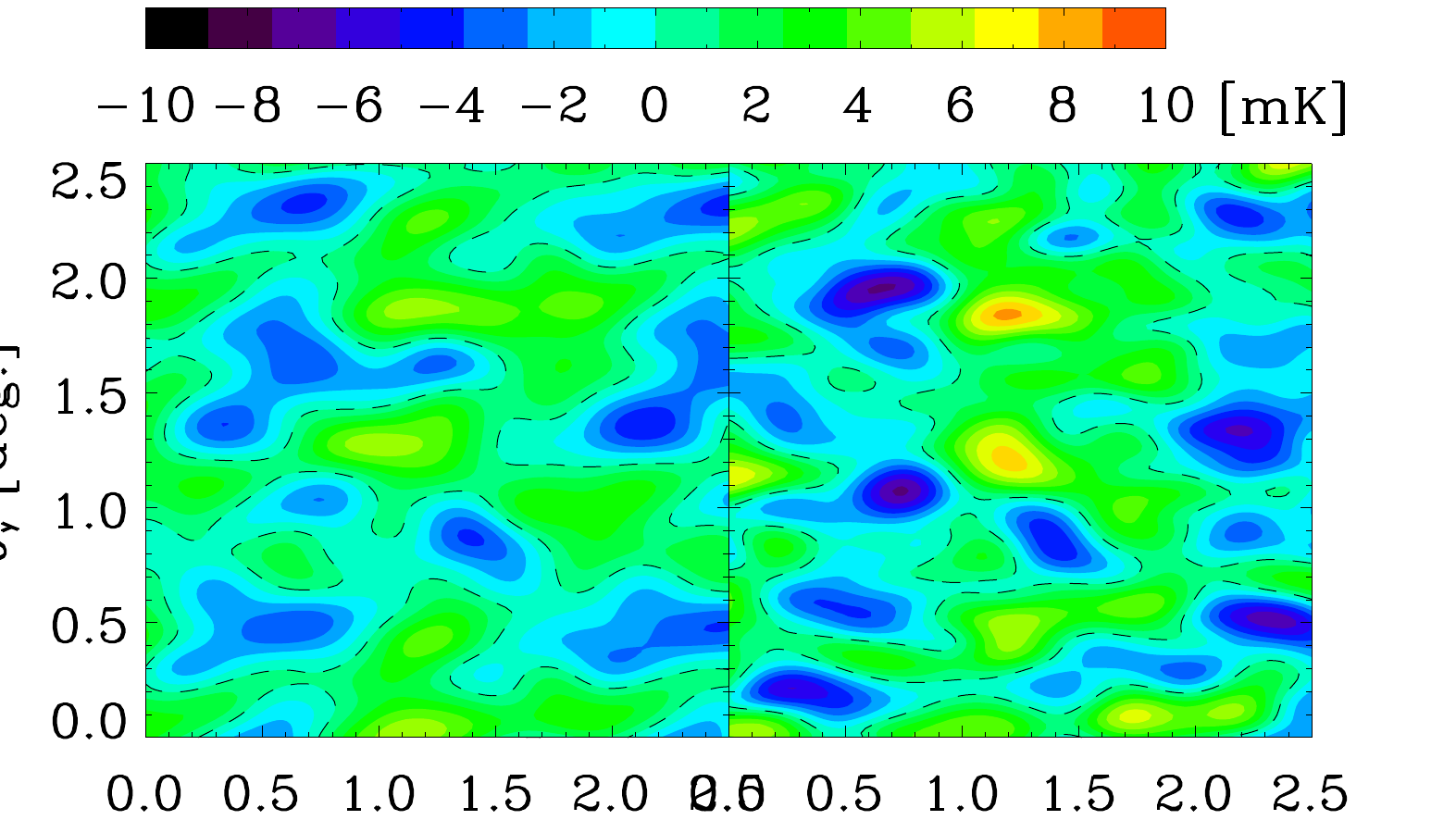}
\vspace{.2cm}
\caption{EoR maps at redshift 7 produced by the BEARS algorithm, with a $\langle x_{HI} \rangle=0.5$ field of view of  $2.5^\circ\times2.5^\circ$.
The left panel shows the original 20 arcmin. smoothed simulation
assuming 2400 hours of Observation with LOFAR. The right panel shows the extracted image after adding noise and foregrounds
also smoothed with 20 arcmin. Gaussian. The two maps are clearly very similar with a correlation coefficient of $\approx 0.61$.
}
\label{fig:mapBEARS}
\end{figure}

We repeated the same procedure on the $200~\cMpch$ {\tt\small\ BEARS} simulations and get very similar results. To cover the 
same angular size as the previous simulation we tile the BEARS simulation box to reach $400~\cMpch$.
The result of this simulation is shown in Fig.~\ref{fig:mapBEARS}, where the left panel shows the original 20 arcmin. smoothed simulation
assuming 2400 hours of Observation with LOFAR. The right panel shows the extracted image after adding noise and foregrounds
also smoothed with 20 arcmin. Gaussian. The two maps are clearly very similar with a correlation coefficient of $\approx 0.61$.
The correlation coefficient here is lower that the same comparison done with the previous simulation (between panels B and F
in Fig.~\ref{fig:maps}) due the relatively small size of the simulation box where the number of large-scales modes is smaller.

This conclusion is insensitive to the type of source we assume to power reionization, i.e., thermal or power-law.
This is reassuring and indicates that this effect is driven more by the very large-scale structure than by the details of the reionization
process. It should be emphasized here that with higher resolution the increased number of low-mass sources
might slightly change the picture, but not in a drastic way, as already seen in very large-scale high-resolution simulations (Iliev \& Mellema, private 
communication). 

\begin{figure*}
\centering
\hspace{0cm}
\includegraphics[width=0.49\textwidth]{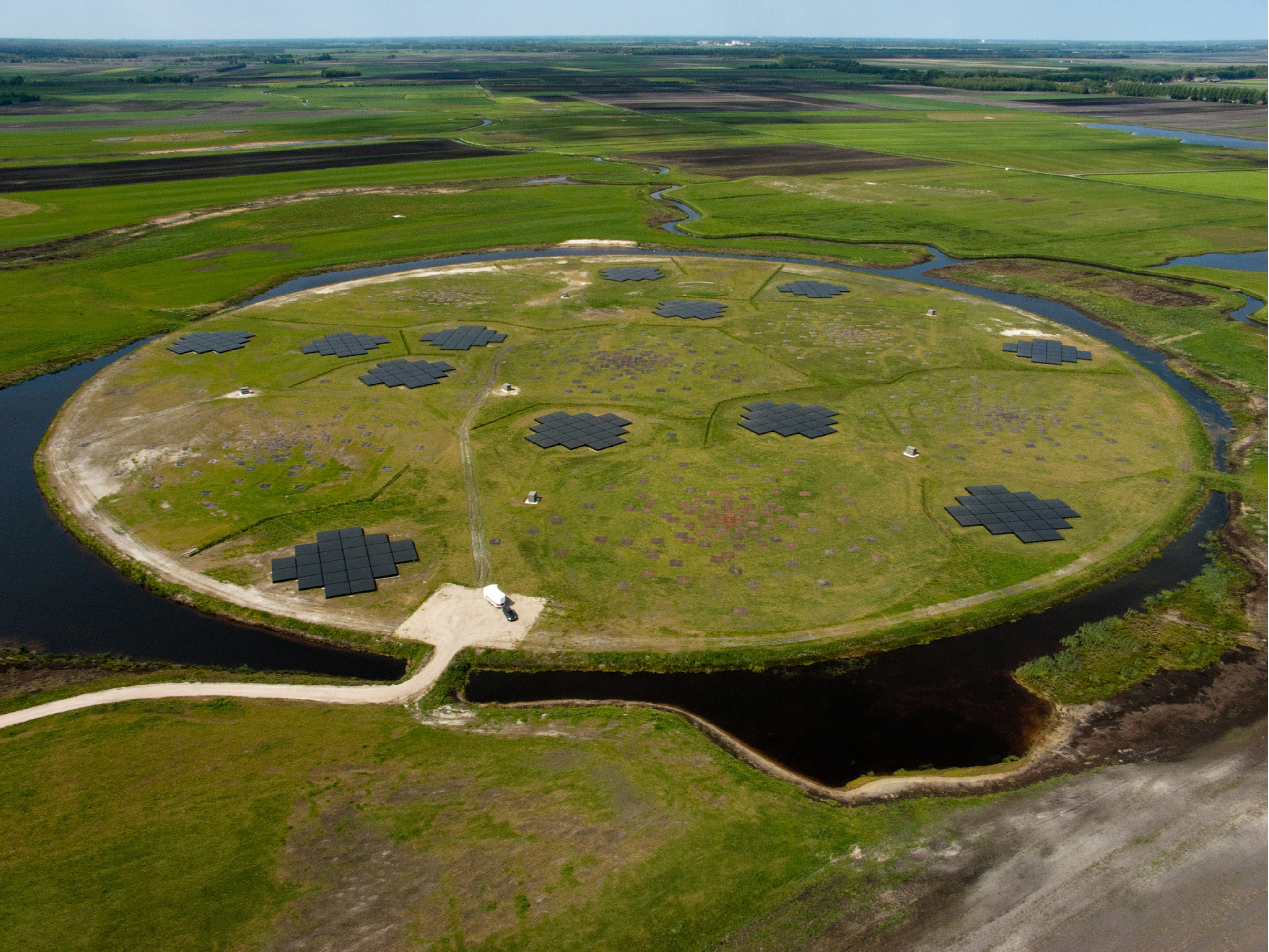}
\hspace{0.3cm}
\includegraphics[width=0.37\textwidth]{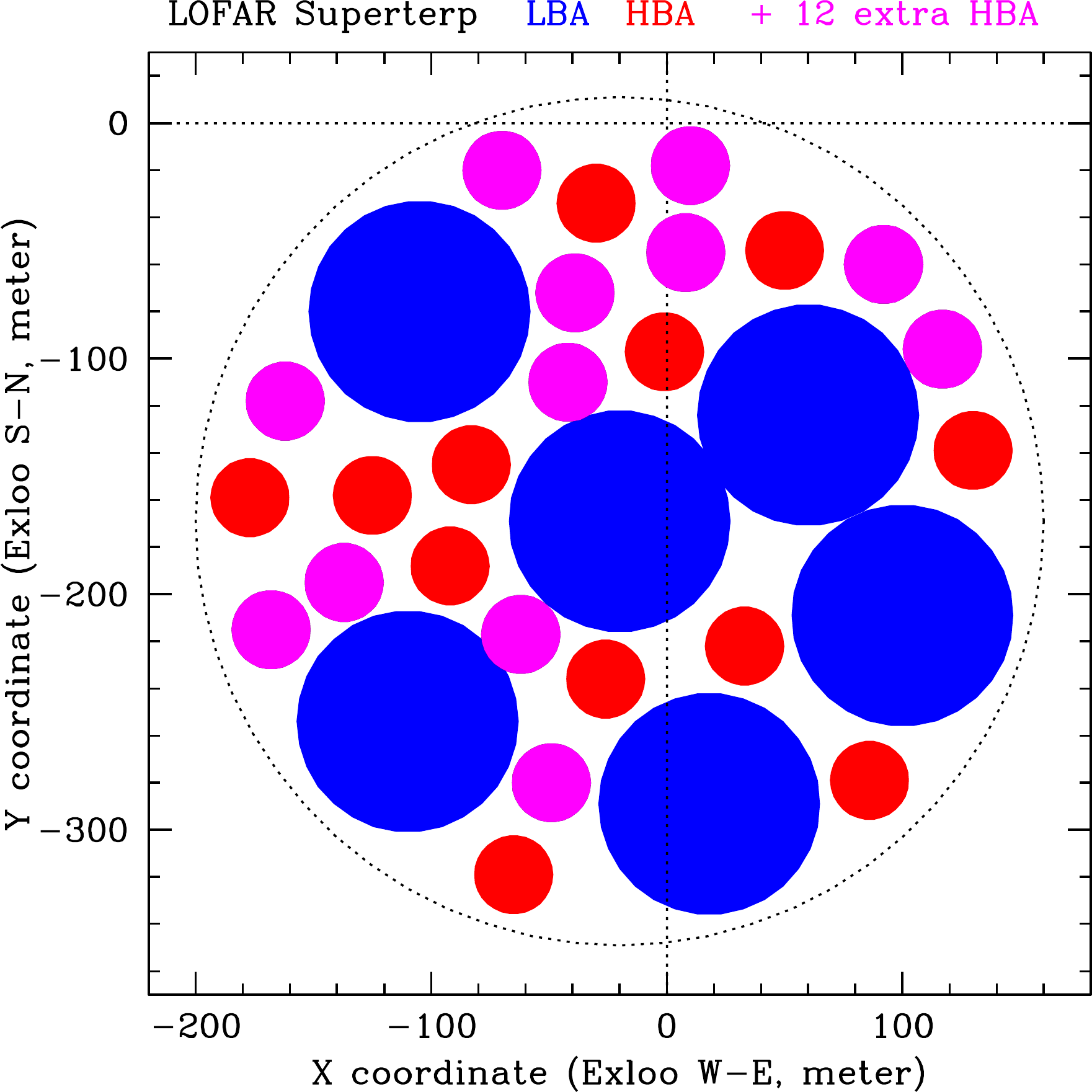}
\caption{{The very central area of the LOFAR core, the ``superterp''. The left panel shows a picture of 
the superterp that was taken in the spring of 2011 (credit for the picture goes to "TopFoto, Assen"). The right side shows a sketch of the layout
of the stations in the superterp. The large blue filled circles indicate the locations of the existing LOFAR 6 Low Band Antenna (LBA) stations 
in the superterp. The red filled circles show the locations of the existing 12 High Band Antenna (HBA) stations, whereas the pink 
filled circles show a possible configuration for an additional 12 HBA stations which would double the collecting area. The sizes of the stations are to scale. The addition of 12 HBA stations 
would significantly increase the sensitivity of LOFAR on large scales.
}}
\label{fig:superterp}
\end{figure*}

\section{Enhancing the Large-Scale Imaging Capabilities of LOFAR}
\label{sec:superterp}

Clearly, the best imaging quality is obtained  when one assumes a very large amount of observing time
(2400 hours) focusing on one single field (Fig.~\ref{fig:maps}). Obviously, this is a vast amount of telescope time that is hard to accumulate especially on open time telescopes, such as LOFAR will become within a number of years. Hence, in what follows we show that a relatively inexpensive modification of the
LOFAR telescope can significantly enhance its sensitivity on large scales.

\begin{figure*}
\centering
\hspace{0cm}
\includegraphics[width=0.95\textwidth]{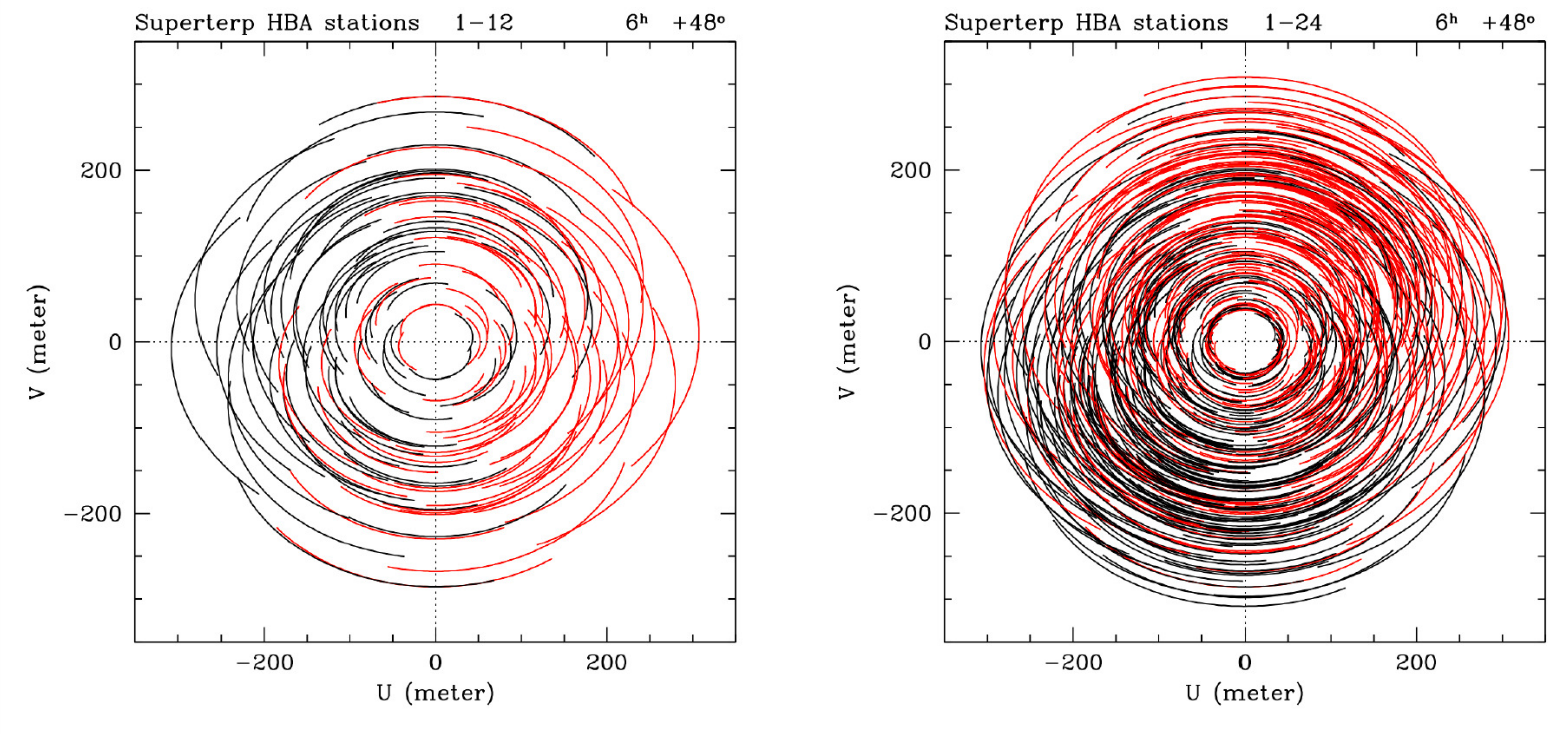}
\vspace{.2cm}
\caption{{The uv-plane covered by the current HBA stations after 6 hours of observation
at a declination of $+48^\circ$. The left panel shows the uv-plane obtained with the current superterp HBA stations whereas the right
panel shows the uv-plane with the proposed additional 12 HBA stations.
}}
\label{fig:uvsuperterp}
\end{figure*}

Increasing the signal-to-noise of the data on scales  $\gsim$ 30 arcmin, corresponding to a comoving scales of $\gsim 60$~ $\cMpch$ at high 
redshifts,  will allow imaging the EoR, especially the last phases of reionization ($\langle x_{HI}\rangle\lsim 0.5$) within a 
very reasonable amount of observational time. 
The quality of imaging on such angular scales for a 2~meter wavelength depends on the number of baselines with length of  $\lsim$200 meters, and 
hence on the number  of stations within the very central  area, i.e., around and within the so-called superterp of LOFAR. For clarity, High Band Antenna 
(HBA) station here means a collection of 24 
HBA tiles with each tile having $4\times 4$ antennas (dipoles). 
The superterp, which in Dutch means ``super mound'', is shown in the left panel of Fig.~\ref{fig:superterp} and is a circular area with a radius of about 
150~m (baselines up to 300~m). In this area there are currently about 12 HBA stations of LOFAR that operate in the 
frequency range of 115- 230~MHz. 
Below we consider the effect of increasing the number of HBA stations in the superterp as a possible way to cut the observation time
to reach the desired sensitivity for imaging.
The right panel of Fig.~\ref{fig:superterp},  shows a possible configuration of 24 stations (pink filled circles)  in the superterp -- the current 12 stations 
(red filled circles) and 12 new stations (pink filled circles). This would increase the number of baselines within the superterp and the area immediately 
surrounding it by a factor of four. The signal-to-noise for the same time of observation would therefore be enhanced by a factor of, at least, two, .i.e., reducing the integration time by a factor of four as well. An addition of HBA stations would not be too 
costly in relative terms since most of the infra structure needed for rolling out the added station will be relatively small.
Fig.~\ref{fig:uvsuperterp} shows the improvement in the uv coverage for the case of a 24 HBA stations in the superterp (right panel) relative to
that of the current 12 HBA stations (left panel). Such an enhancement, which is relatively easy and cheap to obtain, would increase the
sensitivity of LOFAR on these scales by a factor of two, thus boosting the LOFAR imaging capability at large scales. Such an enhancement 
would enable the LOFAR-EoR project to image the reionization process on large-scales rather than only detecting it statistically.
Another issue that we are studying is the possibility to correlate all the tiles within the superterp, instead of whole stations, which will 
add many baselines to the measurement.
This extension proposal has to be studied in more detail before accurate estimations of its ramifications can be appreciated.

\section{Conclusions \& Discussion}
\label{sec:conclusions}

A new generation of low frequency interferometers has recently come on line (LOFAR, MWA, PAPER, SKA). 
Exploring the EoR is a major science driver for these telescopes.
The current common wisdom in the field is that these telescopes will detect the EoR statistically but
that they will not be able to image it, due to their poor sensitivity. The low sensitivity affects both small scales
and large scales. The influence of the noise on small scales is quite clear; its influence on
large scales is indirect and has to do with the ability to fit the foregrounds well along the frequency direction.
Since this fitting is done for each image or uv plane pixel along the frequency direction, the accuracy of the
fit will depend sensitively on the noise level in the data. Hence, the low signal-to-noise will decrease the quality
of the foreground fit, especially on large scale where the foregrounds power becomes larger. With the projected 
level of noise per resolution element in the current generation experiments it is no wonder that imaging has been 
deemed possible only with future instruments such as the Square Kilometer Array \citep[see e.g.,][]{zaroubi10}.

In this paper we have shown that imaging of the neutral IGM at the later stages of reionization
with the current generation of radio interferometers is possible on very large scales ($\gsim 0.5^\circ$). 
The existence of very large-scale neutral regions towards the later phases of the reionization process and their large
contrast with the ionized regions enhances the EoR signal in two ways: firstly, the large neutral regions increase the amplitude
beyond that expected from mere cosmological density fluctuations, secondly, their large size gives a coherent 
signature over about a hundred resolution elements for LOFAR, which overcomes the poor signal-to-noise ratio in each 
of them.  These two effects make it possible in principle to image the reionization process on large scales.

A simple argument in support of our conclusion could also be cast as follows. The ionizating sources
are preferentially located in the high-density regions which typically ionize before the low density regions. Since the density fluctuation power
spectrum peaks at $120~\cMpch$ this will be roughly  the scale of the ionized and neutral regions at 
the midpoint of reionization. It is this scale that essentially allows us to image the EoR on large scales with
instruments like LOFAR. Notice that this argument holds, almost, regardless of the  type of reionization sources.

The second issue we address here, is the issue of the influence of the foregrounds on
the extracted EoR signal, especially on large scales. We show that for realistic foregrounds and noise models 
current state-of-the-art extraction techniques do very well even on scales in excess  of a degree, which
have so far been thought so far to be inaccessible to the current generation of experiments. This again demonstrates 
that imaging of the EoR on large scales is possible with LOFAR.
We have also shown that a modest enhancement of the capabilities of LOFAR at the central area of the core, known as
the superterp, would greatly boost the possibility of imaging the EoR on large scales.

The importance of imaging the EoR with current telescopes cannot be overstated. Astrophysically, imaging would allow
addressing a large number of issues that would otherwise be difficult to deal with. For example, it would
make it possible to identify spatially where the ionized regions are and hence it would allow targeting these regions with follow
up optical and infrared surveys, and iterating on the foregrounds' subtraction \citep{petrovic10}. However, given the issues that face the current experiments, imaging will be
of utmost importance in discovering and addressing systematic effects whose existence would otherwise be impossible to
realize. In other words, imaging of the EoR would boost our confidence in the reliability of the measured signal and
the properties attributed to it. Conversely, if the measured power spectrum indicates the existence of EoR power on 
very large-scales, then the availability of images on such scales would allow us to determine whether this is a result of the cosmological signal or of systematic effects
that have not yet been brought under control.

In the future, the Square Kilometer Array (SKA) will provide enough signal-to-noise to image the EoR
with very high accuracy on scales up to 5$ ^\circ$ and  down to scales of the order of 1 arcmin. This obviously
will surpass LOFAR's performance. SKA will also be able to go to much lower frequencies (down to 50~MHz) enabling a direct
observation of the Universe's Dark Ages. However, SKA will still take about a decade to become operational whereas LOFAR
and the other current instruments are already here, and can make significant scientific discoveries that can be enhanced even further with relatively little extra cost.
 
\section{Acknowledgment}
SZ would like to thank the Lady Davis Foundation and The Netherlands Organisation for Scientific Research (NWO) VICI grant for financial support.
LVEK thanks the European Research Council Starting Grant for support. GH is a member of the LUNAR consortium, which is funded by the NASA Lunar Science Institute (via Cooperative Agreement NNA09DB30A) to investigate concepts for astrophysical observatories on the Moon.


\begin{thebibliography}{60}
 \expandafter\ifx\csname natexlab\endcsname\relax\def\natexlab#1{#1}\fi
 
 \bibitem[{{Baek} {et~al.}(2010){Baek}, {Semelin}, {Di Matteo}, {Revaz}, \&
   {Combes}}]{baek10}
 {Baek} S., {Semelin} B., {Di Matteo} P., {Revaz} Y., {Combes} F., 2010, \aap,
   523, A4
 
 \bibitem[{{Barkana} \& {Loeb}(2005)}]{barkana05}
 {Barkana} R., {Loeb} A., 2005, \apjl, 624, L65
 
 \bibitem[{{Bernardi} {et~al.}(2009){Bernardi}, {de Bruyn}, {Brentjens},
   {Ciardi}, {Harker}, {Jeli{\'c}}, {Koopmans}, {Labropoulos}, {Offringa},
   {Pandey}, {Schaye}, {Thomas}, {Yatawatta}, \& {Zaroubi}}]{bernardi09}
 {Bernardi} G., {de Bruyn} A.~G., {Brentjens} M.~A., {Ciardi} B., {Harker} G.,
   {Jeli{\'c}} V., {Koopmans} L.~V.~E., {Labropoulos} P., {Offringa} A.,
   {Pandey} V.~N., {Schaye} J., {Thomas} R.~M., {Yatawatta} S., {Zaroubi} S.,
   2009, \aap, 500, 965
 
 \bibitem[{{Bernardi} {et~al.}(2010){Bernardi}, {de Bruyn}, {Harker},
   {Brentjens}, {Ciardi}, {Jeli{\'c}}, {Koopmans}, {Labropoulos}, {Offringa},
   {Pandey}, {Schaye}, {Thomas}, {Yatawatta}, \& {Zaroubi}}]{bernardi10}
 {Bernardi} G., {de Bruyn} A.~G., {Harker} G., {Brentjens} M.~A., {Ciardi} B.,
   {Jeli{\'c}} V., {Koopmans} L.~V.~E., {Labropoulos} P., {Offringa} A.,
   {Pandey} V.~N., {Schaye} J., {Thomas} R.~M., {Yatawatta} S., {Zaroubi} S.,
   2010, \aap, 522, A67
 
 \bibitem[{{Bolton} {et~al.}(2010){Bolton}, {Becker}, {Wyithe}, {Haehnelt}, \&
   {Sargent}}]{bolton10}
 {Bolton} J.~S., {Becker} G.~D., {Wyithe} J.~S.~B., {Haehnelt} M.~G., {Sargent}
   W.~L.~W., 2010, \mnras, 771
 
 \bibitem[{{Bolton} \& {Haehnelt}(2007)}]{bolton07}
 {Bolton} J.~S., {Haehnelt} M.~G., 2007, \mnras, 382, 325
 
 \bibitem[{{Bouwens} {et~al.}(2011){Bouwens}, {Illingworth}, {Labbe}, {Oesch},
   {Trenti}, {Carollo}, {van Dokkum}, {Franx}, {Stiavelli}, {Gonz{\'a}lez},
   {Magee}, \& {Bradley}}]{bouwens11}
 {Bouwens} R.~J., {Illingworth} G.~D., {Labbe} I., {Oesch} P.~A., {Trenti} M.,
   {Carollo} C.~M., {van Dokkum} P.~G., {Franx} M., {Stiavelli} M.,
   {Gonz{\'a}lez} V., {Magee} D., {Bradley} L., 2011, \nat, 469, 504
 
 \bibitem[{{Bouwens} {et~al.}(2010){Bouwens}, {Illingworth}, {Oesch},
   {Stiavelli}, {van Dokkum}, {Trenti}, {Magee}, {Labb{\'e}}, {Franx},
   {Carollo}, \& {Gonzalez}}]{bouwens10}
 {Bouwens} R.~J., {Illingworth} G.~D., {Oesch} P.~A., {Stiavelli} M., {van
   Dokkum} P., {Trenti} M., {Magee} D., {Labb{\'e}} I., {Franx} M., {Carollo}
   C.~M., {Gonzalez} V., 2010, \apjl, 709, L133
 
 \bibitem[{{Bowman} {et~al.}(2006){Bowman}, {Morales}, \& {Hewitt}}]{bowman06}
 {Bowman} J.~D., {Morales} M.~F., {Hewitt} J.~N., 2006, \apj, 638, 20
 
 \bibitem[{{Bunker} {et~al.}(2010){Bunker}, {Wilkins}, {Ellis}, {Stark},
   {Lorenzoni}, {Chiu}, {Lacy}, {Jarvis}, \& {Hickey}}]{bunker10}
 {Bunker} A.~J., {Wilkins} S., {Ellis} R.~S., {Stark} D.~P., {Lorenzoni} S.,
   {Chiu} K., {Lacy} M., {Jarvis} M.~J., {Hickey} S., 2010, \mnras, 409, 855
 
 \bibitem[{{Calverley} {et~al.}(2011){Calverley}, {Becker}, {Haehnelt}, \&
   {Bolton}}]{calverley11}
 {Calverley} A.~P., {Becker} G.~D., {Haehnelt} M.~G., {Bolton} J.~S., 2011,
   \mnras, 412, 2543
 
 \bibitem[{{Chapman} {et~al.}(2012){Chapman}, {Abdalla}, {Harker}, {Jeli{\'c}},
   {Labropoulos}, {Zaroubi}, {Brentjens}, {de Bruyn}, \& {Koopmans}}]{chapman12}
 {Chapman} E., {Abdalla} F.~B., {Harker} G., {Jeli{\'c}} V., {Labropoulos} P.,
   {Zaroubi} S., {Brentjens} M.~A., {de Bruyn} A.~G., {Koopmans} L.~V.~E., 2012,
   ArXiv e-prints
 
 \bibitem[{{Ciardi} \& {Madau}(2003)}]{ciardi03a}
 {Ciardi} B., {Madau} P., 2003, \apj, 596, 1
 
 \bibitem[{{Datta} {et~al.}(2010){Datta}, {Bowman}, \& {Carilli}}]{datta10}
 {Datta} A., {Bowman} J.~D., {Carilli} C.~L., 2010, \apj, 724, 526
 
 \bibitem[{{Datta} {et~al.}(2012){Datta}, {Friedrich}, {Mellema}, {Iliev}, \&
   {Shapiro}}]{datta12}
 {Datta} K.~K., {Friedrich} M.~M., {Mellema} G., {Iliev} I.~T., {Shapiro} P.~R.,
   2012, ArXiv e-prints
 
 \bibitem[{{Fan,~\textit{et al.}}(2003)}]{fan03}
 {Fan,~\textit{et al.}} X., 2003, \aj, 125, 1649
 
 \bibitem[{{Fan,~\textit{et al.}}(2006)}]{fan06}
 ---, 2006, \aj, 131, 1203
 
 \bibitem[{{Field}(1958)}]{field58}
 {Field} G.~B., 1958, Proc. IRE, 46, 240
 
 \bibitem[{{Field}(1959)}]{field59b}
 ---, 1959, \apj, 129, 536
 
 \bibitem[{{Furlanetto} {et~al.}(2006){Furlanetto}, {Oh}, \&
   {Briggs}}]{furlanetto06a}
 {Furlanetto} S.~R., {Oh} S.~P., {Briggs} F.~H., 2006, \physrep, 433, 181
 
 \bibitem[{{Geil} {et~al.}(2011){Geil}, {Gaensler}, \& {Wyithe}}]{geil11}
 {Geil} P.~M., {Gaensler} B.~M., {Wyithe} J.~S.~B., 2011, \mnras, 1416
 
 \bibitem[{{Geil} {et~al.}(2008){Geil}, {Wyithe}, {Petrovic}, \& {Oh}}]{geil08}
 {Geil} P.~M., {Wyithe} J.~S.~B., {Petrovic} N., {Oh} S.~P., 2008, \mnras, 390,
   1496
 
 \bibitem[{{Giardino} {et~al.}(2002){Giardino}, {Banday}, {G{\'o}rski},
   {Bennett}, {Jonas}, \& {Tauber}}]{giardino02}
 {Giardino} G., {Banday} A.~J., {G{\'o}rski} K.~M., {Bennett} K., {Jonas} J.~L.,
   {Tauber} J., 2002, \aap, 387, 82
 
 \bibitem[{{Harker} {et~al.}(2010){Harker}, {Zaroubi}, {Bernardi}, {Brentjens},
   {de Bruyn}, {Ciardi}, {Jeli{\'c}}, {Koopmans}, {Labropoulos}, {Mellema},
   {Offringa}, {Pandey}, {Pawlik}, {Schaye}, {Thomas}, \&
   {Yatawatta}}]{harker10}
 {Harker} G., {Zaroubi} S., {Bernardi} G., {Brentjens} M.~A., {de Bruyn} A.~G.,
   {Ciardi} B., {Jeli{\'c}} V., {Koopmans} L.~V.~E., {Labropoulos} P., {Mellema}
   G., {Offringa} A., {Pandey} V.~N., {Pawlik} A.~H., {Schaye} J., {Thomas}
   R.~M., {Yatawatta} S., 2010, \mnras, 405, 2492
 
 \bibitem[{{Harker} {et~al.}(2009{\natexlab{a}}){Harker}, {Zaroubi}, {Bernardi},
   {Brentjens}, {de Bruyn}, {Ciardi}, {Jeli{\'c}}, {Koopmans}, {Labropoulos},
   {Mellema}, {Offringa}, {Pandey}, {Schaye}, {Thomas}, \&
   {Yatawatta}}]{harker09b}
 {Harker} G., {Zaroubi} S., {Bernardi} G., {Brentjens} M.~A., {de Bruyn} A.~G.,
   {Ciardi} B., {Jeli{\'c}} V., {Koopmans} L.~V.~E., {Labropoulos} P., {Mellema}
   G., {Offringa} A., {Pandey} V.~N., {Schaye} J., {Thomas} R.~M., {Yatawatta}
   S., 2009{\natexlab{a}}, \mnras, 397, 1138
 
 \bibitem[{{Harker} {et~al.}(2009{\natexlab{b}}){Harker}, {Zaroubi}, {Thomas},
   {Jeli{\'c}}, {Labropoulos}, {Mellema}, {Iliev}, {Bernardi}, {Brentjens}, {de
   Bruyn}, {Ciardi}, {Koopmans}, {Pandey}, {Pawlik}, {Schaye}, \&
   {Yatawatta}}]{harker09a}
 {Harker} G.~J.~A., {Zaroubi} S., {Thomas} R.~M., {Jeli{\'c}} V., {Labropoulos}
   P., {Mellema} G., {Iliev} I.~T., {Bernardi} G., {Brentjens} M.~A., {de Bruyn}
   A.~G., {Ciardi} B., {Koopmans} L.~V.~E., {Pandey} V.~N., {Pawlik} A.~H.,
   {Schaye} J., {Yatawatta} S., 2009{\natexlab{b}}, \mnras, 393, 1449
 
 \bibitem[{{Iliev} {et~al.}(2008){Iliev}, {Mellema}, {Pen}, {Bond}, \&
   {Shapiro}}]{iliev08}
 {Iliev} I.~T., {Mellema} G., {Pen} U., {Bond} J.~R., {Shapiro} P.~R., 2008,
   \mnras, 384, 863
 
 \bibitem[{{Jeli{\'c}} {et~al.}(2010){Jeli{\'c}}, {Zaroubi}, {Labropoulos},
   {Bernardi}, {de Bruyn}, \& {Koopmans}}]{jelic10b}
 {Jeli{\'c}} V., {Zaroubi} S., {Labropoulos} P., {Bernardi} G., {de Bruyn}
   A.~G., {Koopmans} L.~V.~E., 2010, \mnras, 409, 1647
 
 \bibitem[{{Jeli{\'c}} {et~al.}(2008){Jeli{\'c}}, {Zaroubi}, {Labropoulos},
   {Thomas}, {Bernardi}, {Brentjens}, {de Bruyn}, {Ciardi}, {Harker},
   {Koopmans}, b~{Pandey}, {Schaye}, \& {Yatawatta}}]{jelic08}
 {Jeli{\'c}} V., {Zaroubi} S., {Labropoulos} P., {Thomas} R.~M., {Bernardi} G.,
   {Brentjens} M.~A., {de Bruyn} A.~G., {Ciardi} B., {Harker} G., {Koopmans}
   L.~V.~E., b~{Pandey} V.~N., {Schaye} J., {Yatawatta} S., 2008, \mnras, 389,
   1319
 
 \bibitem[{{Kazemi} {et~al.}(2011){Kazemi}, {Yatawatta}, {Zaroubi},
   {Lampropoulos}, {de Bruyn}, {Koopmans}, \& {Noordam}}]{kazemi11}
 {Kazemi} S., {Yatawatta} S., {Zaroubi} S., {Lampropoulos} P., {de Bruyn} A.~G.,
   {Koopmans} L.~V.~E., {Noordam} J., 2011, \mnras, 414, 1656
 
 \bibitem[{{Labropoulos} {et~al.}(2009){Labropoulos}, {Koopmans}, {Jelic},
   {Yatawatta}, {Thomas}, {Bernardi}, {Brentjens}, {de Bruyn}, {Ciardi},
   {Harker}, {Offringa}, {Pandey}, {Schaye}, \& {Zaroubi}}]{panos09}
 {Labropoulos} P., {Koopmans} L.~V.~E., {Jelic} V., {Yatawatta} S., {Thomas}
   R.~M., {Bernardi} G., {Brentjens} M., {de Bruyn} G., {Ciardi} B., {Harker}
   G., {Offringa} A., {Pandey} V.~N., {Schaye} J., {Zaroubi} S., 2009, ArXiv
   e-prints
 
 \bibitem[{{M\"achler}(1995)}]{machler95}
 {M\"achler} M., 1995, Annals of Statistics, 23, 1496
 
 \bibitem[{{Madau} {et~al.}(1997){Madau}, {Meiksin}, \& {Rees}}]{madau97}
 {Madau} P., {Meiksin} A., {Rees} M.~J., 1997, \apj, 475, 429
 
 \bibitem[{{McQuinn} {et~al.}(2006){McQuinn}, {Zahn}, {Zaldarriaga},
   {Hernquist}, \& {Furlanetto}}]{mcquinn06}
 {McQuinn} M., {Zahn} O., {Zaldarriaga} M., {Hernquist} L., {Furlanetto} S.~R.,
   2006, \apj, 653, 815
 
 \bibitem[{{Mesinger} \& {Furlanetto}(2007)}]{mesinger07}
 {Mesinger} A., {Furlanetto} S., 2007, \apj, 669, 663
 
 \bibitem[{{Mesinger} {et~al.}(2010){Mesinger}, {Furlanetto}, \&
   {Cen}}]{mesinger10}
 {Mesinger} A., {Furlanetto} S., {Cen} R., 2010, ArXiv e-prints
 
 \bibitem[{{Morales} \& {Hewitt}(2004)}]{morales04}
 {Morales} M.~F., {Hewitt} J., 2004, \apj, 615, 7
 
 \bibitem[{{Oesch} {et~al.}(2010){Oesch}, {Bouwens}, {Illingworth}, {Carollo},
   {Franx}, {Labb{\'e}}, {Magee}, {Stiavelli}, {Trenti}, \& {van
   Dokkum}}]{oesch10}
 {Oesch} P.~A., {Bouwens} R.~J., {Illingworth} G.~D., {Carollo} C.~M., {Franx}
   M., {Labb{\'e}} I., {Magee} D., {Stiavelli} M., {Trenti} M., {van Dokkum}
   P.~G., 2010, \apjl, 709, L16
 
 \bibitem[{{Page} {et~al.}(2007){Page}, {Hinshaw}, {Komatsu}, {Nolta},
   {Spergel}, {Bennett}, {Barnes}, {Bean}, {Dor{\'e}}, {Dunkley}, {Halpern},
   {Hill}, {Jarosik}, {Kogut}, {Limon}, {Meyer}, {Odegard}, {Peiris}, {Tucker},
   {Verde}, {Weiland}, {Wollack}, \& {Wright}}]{page07}
 {Page} L., {Hinshaw} G., {Komatsu} E., {Nolta} M.~R., {Spergel} D.~N.,
   {Bennett} C.~L., {Barnes} C., {Bean} R., {Dor{\'e}} O., {Dunkley} J.,
   {Halpern} M., {Hill} R.~S., {Jarosik} N., {Kogut} A., {Limon} M., {Meyer}
   S.~S., {Odegard} N., {Peiris} H.~V., {Tucker} G.~S., {Verde} L., {Weiland}
   J.~L., {Wollack} E., {Wright} E.~L., 2007, \apjs, 170, 335
 
 \bibitem[{{Petrovic} \& {Oh}(2010)}]{petrovic10}
 {Petrovic} N., {Oh} S.~P., 2010, ArXiv e-prints
 
 \bibitem[{{Pritchard} \& {Furlanetto}(2007)}]{pritchard07}
 {Pritchard} J.~R., {Furlanetto} S.~R., 2007, \mnras, 376, 1680
 
 \bibitem[{{Pritchard} \& {Loeb}(2008)}]{pritchard08}
 {Pritchard} J.~R., {Loeb} A., 2008, \prd, 78, 103511
 
 \bibitem[{{Pritchard} \& {Loeb}(2010)}]{pritchard10}
 ---, 2010, \prd, 82, 023006
 
 \bibitem[{{Pritchard} \& {Loeb}(2011)}]{pritchard12}
 ---, 2011, ArXiv e-prints
 
 \bibitem[{{Rybicki} \& {Lightman}(1986)}]{rybicki86}
 {Rybicki} G.~B., {Lightman} A.~P., 1986, {Radiative Processes in Astrophysics},
   Rybicki G.~B.~\&~Lightman A.~P., ed.
 
 \bibitem[{{Santos} {et~al.}(2005){Santos}, {Cooray}, \& {Knox}}]{santos05}
 {Santos} M.~G., {Cooray} A., {Knox} L., 2005, \apj, 625, 575
 
 \bibitem[{{Santos} {et~al.}(2010){Santos}, {Ferramacho}, {Silva}, {Amblard}, \&
   {Cooray}}]{santos10}
 {Santos} M.~G., {Ferramacho} L., {Silva} M.~B., {Amblard} A., {Cooray} A.,
  2010, \mnras, 406, 2421

 \bibitem[{{Shaver} {et~al.}(1999){Shaver}, {Windhorst}, {Madau}, \& {de
   Bruyn}}]{shaver99}
 {Shaver} P.~A., {Windhorst} R.~A., {Madau} P., {de Bruyn} A.~G., 1999, \aap,
   345, 380
 
 \bibitem[{{Spergel} {et~al.}(2007){Spergel}, {Bean}, {Dor{\'e}}, {Nolta},
   {Bennett}, {Dunkley}, {Hinshaw}, {Jarosik}, {Komatsu}, {Page}, {Peiris},
   {Verde}, {Halpern}, {Hill}, {Kogut}, {Limon}, {Meyer}, {Odegard}, {Tucker},
   {Weiland}, {Wollack}, \& {Wright}}]{spergel07}
 {Spergel} D.~N., {Bean} R., {Dor{\'e}} O., {Nolta} M.~R., {Bennett} C.~L.,
   {Dunkley} J., {Hinshaw} G., {Jarosik} N., {Komatsu} E., {Page} L., {Peiris}
   H.~V., {Verde} L., {Halpern} M., {Hill} R.~S., {Kogut} A., {Limon} M.,
   {Meyer} S.~S., {Odegard} N., {Tucker} G.~S., {Weiland} J.~L., {Wollack} E.,
   {Wright} E.~L., 2007, \apjs, 170, 377
 
 \bibitem[{{Taylor} {et~al.}(1999){Taylor}, {Carilli}, \& {Perley}}]{taylor99}
 {Taylor} G.~B., {Carilli} C.~L., {Perley} R.~A., eds., 1999, Astronomical
   Society of the Pacific Conference Series, Vol. 180, {Synthesis Imaging in
   Radio Astronomy II}
 
 \bibitem[{{Tegmark} {et~al.}(2000){Tegmark}, {Eisenstein}, {Hu}, \& {de
   Oliveira-Costa}}]{tegmark00}
 {Tegmark} M., {Eisenstein} D.~J., {Hu} W., {de Oliveira-Costa} A., 2000, \apj,
   530, 133
 
 \bibitem[{{Theuns} {et~al.}(2002){Theuns}, {Schaye}, {Zaroubi}, {Kim},
   {Tzanavaris}, \& {Carswell}}]{theuns02}
 {Theuns} T., {Schaye} J., {Zaroubi} S., {Kim} T., {Tzanavaris} P., {Carswell}
   B., 2002, \apjl, 567, L103
 
 \bibitem[{{Thomas} \& {Zaroubi}(2008)}]{thomas08}
 {Thomas} R.~M., {Zaroubi} S., 2008, \mnras, 384, 1080
 
 \bibitem[{{Thomas} \& {Zaroubi}(2011)}]{thomas11}
 ---, 2011, \mnras, 410, 1377
 
 \bibitem[{{Thomas} {et~al.}(2009){Thomas}, {Zaroubi}, {Ciardi}, {Pawlik},
   {Labropoulos}, {Jeli{\'c}}, {Bernardi}, {Brentjens}, {de Bruyn}, {Harker},
   {Koopmans}, {Mellema}, {Pandey}, {Schaye}, \& {Yatawatta}}]{thomas09}
 {Thomas} R.~M., {Zaroubi} S., {Ciardi} B., {Pawlik} A.~H., {Labropoulos} P.,
   {Jeli{\'c}} V., {Bernardi} G., {Brentjens} M.~A., {de Bruyn} A.~G., {Harker}
   G.~J.~A., {Koopmans} L.~V.~E., {Mellema} G., {Pandey} V.~N., {Schaye} J.,
   {Yatawatta} S., 2009, \mnras, 393, 32
 
 \bibitem[{{Thompson} {et~al.}(2001){Thompson}, {Moran}, \&
   {Swenson}}]{thompson01}
 {Thompson} A.~R., {Moran} J.~M., {Swenson} Jr. G.~W., 2001, {Interferometry and
   Synthesis in Radio Astronomy, 2nd Edition}, {Thompson, A.~R., Moran, J.~M.,
   \& Swenson, G.~W., Jr.}, ed.
 
 \bibitem[{{Yatawatta} {et~al.}(2009){Yatawatta}, {Zaroubi}, {de Bruyn},
   {Koopmans}, \& {Noordam}}]{yatawatta09}
 {Yatawatta} S., {Zaroubi} S., {de Bruyn} G., {Koopmans} L., {Noordam} J., 2009,
   in Digital Signal Processing Workshop and 5th IEEE Signal Processing
   Education Workshop, 2009. DSP/SPE 2009. IEEE 13th
 
 \bibitem[{{Zahn} {et~al.}(2007){Zahn}, {Lidz}, {McQuinn}, {Dutta}, {Hernquist},
   {Zaldarriaga}, \& {Furlanetto}}]{zahn07}
 {Zahn} O., {Lidz} A., {McQuinn} M., {Dutta} S., {Hernquist} L., {Zaldarriaga}
   M., {Furlanetto} S.~R., 2007, \apj, 654, 12
 
 \bibitem[{{Zahn} {et~al.}(2011){Zahn}, {Reichardt}, {Shaw}, {Lidz}, {Aird},
   {Benson}, {Bleem}, {Carlstrom}, {Chang}, {Cho}, {Crawford}, {Crites}, {de
   Haan}, {Dobbs}, {Dore}, {Dudley}, {George}, {Halverson}, {Holder},
   {Holzapfel}, {Hoover}, {Hou}, {Hrubes}, {Joy}, {Keisler}, {Knox}, {Lee},
   {Leitch}, {Lueker}, {Luong-Van}, {McMahon}, {Mehl}, {Meyer}, {Millea},
   {Mohr}, {Montroy}, {Natoli}, {Padin}, {Plagge}, {Pryke}, {Ruhl}, {Schaffer},
   {Shirokoff}, {Spieler}, {Staniszewski}, {Stark}, {Story}, {van Engelen},
   {Vanderlinde}, {Vieira}, \& {Williamson}}]{zahn12}
 {Zahn} O., {Reichardt} C.~L., {Shaw} L., {Lidz} A., {Aird} K.~A., {Benson}
   B.~A., {Bleem} L.~E., {Carlstrom} J.~E., {Chang} C.~L., {Cho} H.~M.,
   {Crawford} T.~M., {Crites} A.~T., {de Haan} T., {Dobbs} M.~A., {Dore} O.,
   {Dudley} J., {George} E.~M., {Halverson} N.~W., {Holder} G.~P., {Holzapfel}
   W.~L., {Hoover} S., {Hou} Z., {Hrubes} J.~D., {Joy} M., {Keisler} R., {Knox}
   L., {Lee} A.~T., {Leitch} E.~M., {Lueker} M., {Luong-Van} D., {McMahon}
   J.~J., {Mehl} J., {Meyer} S.~S., {Millea} M., {Mohr} J.~J., {Montroy} T.~E.,
   {Natoli} T., {Padin} S., {Plagge} T., {Pryke} C., {Ruhl} J.~E., {Schaffer}
   K.~K., {Shirokoff} E., {Spieler} H.~G., {Staniszewski} Z., {Stark} A.~A.,
   {Story} K., {van Engelen} A., {Vanderlinde} K., {Vieira} J.~D., {Williamson}
   R., 2011, ArXiv e-prints
 
 \bibitem[{{Zaroubi}(2010)}]{zaroubi10}
 {Zaroubi} S., 2010, in Widefield Science and Technology for the SKA, {S.A.
   Torchinsky, A. van Ardenne, T. van den Brink-Havinga, A. van Es, A.J.
   Faulkner}, ed., p.~75
 
 \bibitem[{{Zaroubi} {et~al.}(2007){Zaroubi}, {Thomas}, {Sugiyama}, \&
   {Silk}}]{zaroubi07}
 {Zaroubi} S., {Thomas} R.~M., {Sugiyama} N., {Silk} J., 2007, \mnras, 375, 1269
 
 \end{thebibliography}

\end{document}